\documentclass[11pt,a4paper]{article}
\pdfoutput=1
\usepackage{jinstpub1}
\usepackage{lineno}

\begin{document}
%\linenumbers

\def\ge76      {$^{76}$Ge}
\def\tl208      {$^{208}$Tl}
\def\th232      {$^{232}$Th}
\def\th228      {$^{228}$Th}
\def\co60      {$^{60}$Co}
\def\rn222      {$^{222}$Rn}
\def\Qbb       {Q$_{\beta\beta}$}
\def\dbd       {double-beta decay}
\def\znuebb    {$0\nu\beta\beta$}
\def\znubb    {$0\nu\beta\beta$}
\def\tnubb    {$2\nu\beta\beta$}

\def\ONBB      {{$0\nu\beta\beta$}}
\def\onbb      {{$0\nu\beta\beta$}}
\def\ONBBD      {{$0\nu\beta\beta$ decay}}
\def\MBB      {{$m_{\beta\beta}$}}
\def\lps       {$\ell$\,/\,s}
\def\gesix      {{$^{76}$Ge}}
\def\cosixt      {{$^{60}$Co}}
\def\cum        {{m$^3$}}
\def\radtts     {{$^{226}$Ra}}
\def\thtte     {{$^{228}$Th}}
\def\thzdz     {{$^{232}$Th}}
\def\tltoe      {{$^{208}$Tl}}
\def\kfty        {{$^{40}$K}}
\def\gam        {{$\gamma$}}
\def\ln         {LN2}
\def\LN2         {LN2}
\def\lar        {LAr}
\def\etal        {et al.,}
\def\kgy        {kg$\cdot$yr}
\def\ctsper     {cts/(keV$\cdot$kg$\cdot$yr)}
\def\ctsperee     {cts/(keV$_{ee}\cdot$kg$\cdot$yr)}
\def\ctsperrec     {cts/(keV$_{rec}\cdot$kg$\cdot$yr)}
\def\zctsper     {{$10^{-2}$~cts/(keV$\cdot$kg$\cdot$yr)}}
\def\dctsper     {{$10^{-3}$~cts/(keV$\cdot$kg$\cdot$yr)}}
\def\vctsper     {{$10^{-4}$~cts/(keV$\cdot$kg$\cdot$yr)}}
\def\sctsper     {{$10^{-6}$~cts/(keV$\cdot$kg$\cdot$yr)}}
\def\ctsperx     {$\frac{10^{-3}\rm cts}{\rm keV\cdot kg \cdot yr}$}
\def\deg         {$^\circ$C}
\def\hallA       {hall A}

\def\CUORI      {{\mbox{{\sc Cuoricino}}}}
\def\GEANT      {{\mbox{{\sc Geant}}}}
\def\GEM      {{\mbox{{\sc Gem}}}}
\def\GENI      {{\mbox{{\sc Genius}}}}
\def\GERDA       {\mbox{{\sc Gerda}}}
\def\HDM       {\mbox{{\sc HdM}}}
\def\IGEX      {{\mbox{{\sc Igex}}}}
\def\INFN      {{\mbox{{\sc Infn}}}}
\def\LNGS      {{\mbox{{\sc Lngs}}}}
\def\LVD      {{\mbox{{\sc Lvd}}}}
\def\legend   {{\mbox{{\sc Legend-200}}}}
\def\LEGEND   {{\mbox{{\sc Legend}}}}
\def\Majorana      {{\mbox{{\sc Majorana}}}}
\def\NEMO      {{\mbox{{\sc Nemo3}}}}
\def\REG {\textsuperscript{\textregistered}}

\def\GERDA   {\mbox{{\sc Gerda}}}

\title{
Design and Performance of the GERDA \ Low-Background Cryostat for Operation in Water
}

%\author[a]{K. T. Kn\"opfle}%\corref{cor1}}
\author{K. T. Kn\"opfle}%\corref{cor1}}
%\ead{ktkno@mpi-hd.mpg.de}
%\author[a]{and B. Schwingenheuer}
\author{and B. Schwingenheuer}

\affiliation{
Max-Planck-Institut f\"ur Kernphysik, \\ Saupfercheckweg 1, D-69117 Heidelberg, Germany
}

\emailAdd{Karl-Tasso.Knoepfle@mpi-hd.mpg.de}
%\cortext[cor1]{Corresponding author, Tel.: +49 6221 516509; fax +49 6221 516603.}

%%%%%%%%%%%%%%%%%%%%%%%%%%%%%%%%%%%%%%%%%%%%%%%%%%%%%%%%%%%%%%%%%%%%%%%%%%%%%%%%%%%%%%%%%%%%%%%%%
\abstract{
In searching for the neutrinoless double-beta decay of \gesix \
the GERmanium Detector Array (\GERDA ) experiment at the \INFN \ Laboratori 
Nazionali del Gran Sasso has achieved an unprecedented low background of well below \dctsper \  
in the  
region of interest. It has taken advantage of the first realization of a novel 
shielding concept based on a large cryostat filled with a liquid noble gas 
that is immersed in a water tank. The germanium detectors are operated without 
encapsulation in liquid argon. Argon and water shield the environmental 
background from the laboratory and the cryostat construction materials to a negligible level.
The same approach has been adopted in the meantime by various experiments.  
This paper provides an overview of the design and the operation experience of the 64\,m$^3$ 
liquid argon cryostat and its associated infrastructure. The discussion inludes 
the challenging safety issues associated with the operation of a large cryostat 
in a water tank.
}
%%%%%%%%%%%%%%%%%%%%%%%%%%%%%%%%%%%%%%%%%%%%%%%%%%%%%%%%%%%%%%%%%%%%%%%%%%%%%%%%%%%%%%%%%%%%%%%%%

%%%%%%%%%%%%%%%%%%%%%%%%%%%%%%%%%%%%%%%%%%%%%%%%%%%%%%%%%%%%%%%%%%%%%%%%%%%%%%%%%%%%%%%%%%%%%%%%%
\keywords{ 
Double-beta decay detectors, detector design and construction technologies
and materials, overall mechanics design, cryogenics. \hfill 
}
%%%%%%%%%%%%%%%%%%%%%%%%%%%%%%%%%%%%%%%%%%%%%%%%%%%%%%%%%%%%%%%%%%%%%%%%%%%%%%%%%%%%%%%%%%%%%%%%

%\begin{document)

\maketitle

%\flushbottom
%\linenumbers
  
%%%%%%%%%%%%%%%%%%%%%%%%%%%%%%%%%%%%%%%%%%%%%%%%%%%%%%%%%%%%%%%%%%%%%%%%%%%%%%%%%%%%%%%%%%%%%%%%%

\section{Introduction}

The GERmanium Detector Array (\GERDA) collaboration has terminated in February 2020
its search for neutrinoless double-beta (\znubb ) 
decay of \ge76 , \ge76 $\rightarrow$ $^{76}$Se\,+\,2e$^-$ \cite{gphInim,gphIprl,gphIIup,gphIIr1,gphIIr2,gphIIr3,gphIIr4}.
Located
in \hallA \ of the \INFN\ deep-underground Laboratori Nazionali del Gran Sasso (\LNGS ), Italy, 
the experiment used germanium (Ge) diodes fabricated from high purity Ge material enriched in the \gesix \ isotope 
fraction, simultaneously as source and detector. 
The experimental signature for \znuebb\ decay is the observation of a peak in the energy spectrum of 
the 2e$^-$ final state at the  endpoint of the continuous energy spectrum of the standard \tnubb \ double
beta decay, \ge76 $\rightarrow$ $^{76}$Se\,+\,2e$^-$\,+\,2$\bar \nu$, which for \ge76 \ is at the transition 
energy \Qbb  =\,2039\,keV.
The observation of \znubb \ decay 
would have significant implications on particle physics and cosmology: it 
would establish lepton number violation, and the neutrino to be its own anti-particle (Majorana particle)
supporting in this way theoretical explanations of the baryon asymmetry in our universe \cite{baryasym}.

When the \GERDA \ experiment was conceived in 2004 \cite{LOI} previous experiments had shown the detection of this 
hypothetical process to be extremely demanding. Almost 4 decades after the pioneering investigation into 
the \znuebb\ decay of \ge76 \ by the Milano group \cite{fiorini} its half-life limit had been improved by 5 orders 
of magnitude to about 10$^{25}$ years.
Major progress was due to the continuous reduction of the background index (BI) which is the number of events 
at \Qbb \ normalized 
to a 1~keV energy interval and exposure (product of detector mass $M$ and measurement time $t$).  
Another big improvement is due to the use of enriched germanium since the signal-to-background ratio scales with the 
enrichment fraction.
However, none of the experiments was \lq background-free', i.e., had an expected background count  $<1$ at full exposure 
within an energy interval given by the energy resolution around \Qbb.
Hence their lifetime limit improved only as $\sqrt{M\cdot t}$ and not linearly in $(M\cdot t)$ as 
in the \lq background-free' case. It was 
the goal of \GERDA\ to realize for the first time a \lq background-free' experiment \cite{LOI}.
  
\GERDA \ achieved this goal in two steps.
In Phase I (November 2011 to May 2013), it improved the BI at \Qbb \ to \zctsper , a factor of 10 lower than in previous 
Ge experiments \cite{gphIprl}. 
After an upgrade  \cite{gphIIup}, Phase II (December 2015 to May 2018 and July 2018 to December 2019) yielded 
a further reduction of the BI by more than a factor of 10 \cite{gphIIr1,gphIIr2,gphIIr3,gphIIr4}. 
A value  well below \dctsper \  had never been reached so far by any other \znubb \ decay experiment if
normalized by the energy resolution \cite{gphIIr4}.
 
Prerequisite for this success was the transition from the traditional compact lead-copper shielding approach
to the realization of a novel shielding concept. It emerged from the proposal \cite{Heus} to operate Ge detectors 
in ultra-pure liquid nitrogen (\ln ) because of the radiopurity of \ln\ and its low $Z$. As we will discuss
in the next section, \ln \  was replaced by liquid argon (LAr).

\begin{figure} [h]
\begin{center}
\includegraphics[width=0.9\textwidth] {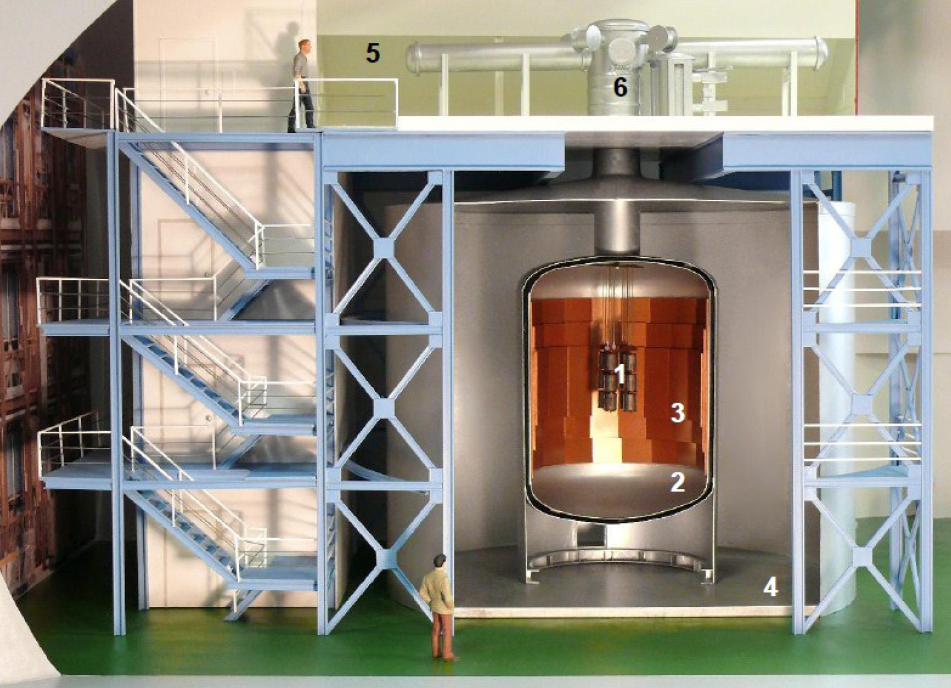}
\caption{Artist's view of the \GERDA \ experiment in \hallA \ of \LNGS\ showing the (enlarged) germanium
detector array (1), the LAr cryostat (2) with internal copper shield (3), the surrounding water tank (4),
the clean room (5), and the lock (6) through which the Ge detectors are deployed into the cryostat.
}
\label{fig:gerda-setup}
\end{center}
\end{figure}

Figure~\ref{fig:gerda-setup} shows the \GERDA\ detector array of bare Ge diodes immersed in a large
%%%%%%%%%%%%%%%%%%%%%%%%%%%%%%%%%%%%%%%%%%%%%%%%%%%%%%%%%%%%%%%%%%%%%%%%%%%%%%%%%%%%%%%%%%%%%%%%
%%%%%%%%%%%%%%%%%%%%%%%%%%%%%%%%%%%%%%%%%%%%%%%%%%%%%%%%%%%%%%%%%%%%%%%%%%%%%%%%%%%%%%%%%%%%%%%% 
volume (64\,m$^3$) of high purity (N5.0) LAr, which serves both as cooling and shielding 
medium. For Phase II, the LAr was turned into an active veto system by adding instrumentation for the 
readout of its scintillation light. 
The LAr is contained in a vacuum-insulated cryostat of 4.2\,m diameter which itself is submerged
in a large (650\,m$^3$) water tank of 10\,m diameter.\footnote{A similar setup with \LN2 \ as cryoliquid 
had been proposed earlier \cite{Zdes}, but has never been realized.}    
The  purified water ($>$\,0.17\,M$\Omega\cdot$m) 
complements the shielding against the radiation from the surrounding rock and concrete; it serves also 
as neutron shield and as the medium for an active muon veto system \cite{muonveto}.

This installation comprises the first and so far the largest cryostat that is operated underground within a 
large water tank. 
The same approach has been adopted in the meantime by various experiments \cite{LUX, XENON, DARKSIDE, PandaX}.
The new scenario of a cryostat immersed in water required to construct
a cryostat of extreme reliability. Because the surrounding water represents a huge heat source, the 
failure/leakage of one of the two or both cryostat shells could result in a huge evaporation or even a
rapid phase transition of the cryogenic liquid 
with potentially severe consequences. Special design features, extensive safety reviews and a first 
measurement of the heat transfer in LAr contributed to a successful mitigation of the associated risks. 
Since its first LAr fill in 2009 the cryostat has been operated without any safety problem.
Thus a major part of the \GERDA \ experimental setup including cryostat and water tank will be used for the
upcoming \legend \ experiment \cite{LEGEND}. 

This paper describes construction and performance of the cryostat including its cryogenic system.
Section 2 discusses the specific design criteria of the water tank - cryostat system and 
the associated safety aspects. 
Section 3 presents the engineering description of the cryostat, 
section 4 the supporting cryogenic infrastructure. 
Section 5 addresses special safety aspects of the operation of
a cryostat in a water tank. Appendix A provides supplementary material on heat transfer in the 
pool boiling regime. 
Section 6 summarizes acceptance tests of the cryostat and its performance. 
Section 7 concludes the paper. Appendix B shows the timeline of the \GERDA \ experiment in more
detail.
   
%%%%%%%%%%%%%%%%%%%%%%%%%%%%%%%%%%%%%%%%%%%%%%%%%%%%%%%%%%%%%%%%%%%%%%%%%%%%%%%%%%%%%%%%%%%%%%%%%%%%%
%%%%%%%%%%%%%%%%%%%%%%%%%%%%%%%%%%%%%%%%%%%%%%%%%%%%%%%%%%%%%%%%%%%%%%%%%%%%%%%%%%%%%%%%%%%%%%%%%
   
\section{Design consideration for the cryostat - water tank system} 
%\subsection{Design criteria}
Two major requirements determined the design of the cryostat and the surrounding water tank, 
(i) reduction of the external \gam \ radiation to the desired level, and (ii) outstanding safe longterm operation.   
The first issue determines the size of cryostat and surrounding water tank as well as the
radio\--purity of construction materials, the second - based on rigorous risk analysis - the layout, 
construction procedures and certification of cryostat and cryogenic infrastructure. Additional
constraints include a minimum water thickness of about 2\,m to moderate neutrons efficiently and to provide
a sufficiently large Cherenkov medium for muon detection. The introduction of the water volume 
allows to use a smaller volume of cryogenic liquid which is highly desirable from safety aspects. 
The diameter of the cryostat is also limited by road transport to less than 4.5~m.  
Last but not least the cryostat - water tank assembly has to fit into the allocated space in \hallA , that is 
within a diameter and height of 10\,m, respectively.  
On top of that, about 2.5\,m in height (4\,m at center) are left for cleanroom and lock through which 
the Ge detectors
can be deployed directly in the \lar \ contained in the cryostat (see Fig.~\ref{fig:gerda-setup} and 
Figs. 2 \& 10 in \cite{gprop}).
    
\subsection{Shielding of external \gam \ background}
The external background consists of \gam \ rays from the primordial decay chains, neutrons and muons.
As to the \gam \ background at \Qbb \ the predominant contribution is expected to be due to the Compton tail of the
2.615\,MeV \gam \ line of the \tl208 \ decay. In \hallA \ of LNGS the respective 2.615 MeV flux (surface activity)
has been measured to be $(0.031\pm 0.09)$cts/(s$\cdot$cm$^2$) \cite{oleg}. Assuming twice this value, measurements 
and Monte Carlo studies have yielded for the BI of a shielded 2\,kg Ge diode in \hallA \ \cite{gprop} 
the approximation of $BI=2250\cdot X\cdot {\rm exp}(-X)$ \ctsper \ 
where $X$\,=\,$\Sigma_i$(t$_i\mu_i$)\,$>$\,5 is the sum of the products of thickness t$_i$ and 
linear absorption coefficient $\mu_i$ of the various shielding materials $i$. 
Table~\ref{tab:att-coeff} lists absorption coefficients and activities of potential shielding materials. 
%%%%%%%%%%%%%%%%%%%%%%%%%%%%%%%%%%%%%%%%%%%%%%%%%%%%%%%%%%%%%%%%%%%%%%%%%%%%%%%%%%%%%%%%%%%%%%%%%%%% 
\renewcommand{\arraystretch}{0.8}
\begin{table}[ht]
%\openup -3\jot
\begin{center}
\caption{Linear attenuation coefficients $\mu$ for 2.615 MeV \gam \ rays in various materials including liquid 
nitrogen (\ln ) and liquid argon (LAr),  
the material's assumed \th228 \ mass activity  A$_{\rm m}$ as well as the corresponding surface activity  A$_{\rm s}$
for thickness t (see \cite{gprop}). Note that, unexpectedly, it was possible to procure stainless steel of significantly
lower radioactivity (see Table 3).
}
\vskip+2truemm
\label{tab:att-coeff}
\begin{tabular}{lcccc}
%\openup -3\jot
\hline
\hline \\[-2.0ex]
 materials &  $\mu$ & $\rho$  & A$_{\rm m}$(\th228 )  &  A$_{\rm s}$(\tl208 )   \\[+0.1ex]
\hline \\[-2.0ex]
               &  [cm$^{-1}$] & [g\,/\,cm$^3$] & [$\mu$Bq/kg] & [$\mu$Bq/cm$^2$]  \\[+0.1ex]
\hline \\[-2.0ex]	       
\ln          &0.0311 & 0.81   & *              & * \\
water       &0.0427 & 1.0    & 1              & t$>$1\,m: 0.01\\
LAr         &0.050  & 1.39   & *              & * \\
steel       &0.299  & 7.87   & 2$\cdot$10$^4$ & t$=$2\,cm: 84.5 \\
Cu          &0.338  & 8.96   & 25             & t$=$3\,cm: 0.15\\
            &       &        &                & t$>$16\,cm: 0.24\\
Pb          &0.484  & 11.35  & 30             & t$>$10\,cm: 0.25 \\
\hline
\hline \\[-2.0ex]
* negligible
\end{tabular}
\end{center}
\end{table}
%%%%%%%%%%%%%%%%%%%%%%%%%%%%%%%%%%%%%%%%%%%%%%%%%%%%%%%%%%%%%%%%%%%%%%%%%%%%%%%%%%%%%%%%%%%%%%%%
The condition $BI$\,=\,\dctsper \ yields $X$\,=\,17.5. It implies that the surface 
activity of \hallA \ of 0.0625~Bq/cm$^2$ has to be reduced by a factor exp(-17.5) to 1.6~nBq/cm$^2$
which yields $t=563$\,cm for \ln .  Hence, to fit into \hallA ,
a graded shield has to be considered in which part of the \ln \ is substituted by materials of 
larger absorption coefficients and adequate radiopurity (see Table~\ref{tab:att-coeff}). Due to 
its high radiopurity water is a perfect substitute for \ln .

\begin{figure} 
\begin{center}
\includegraphics[width=1.0\textwidth] {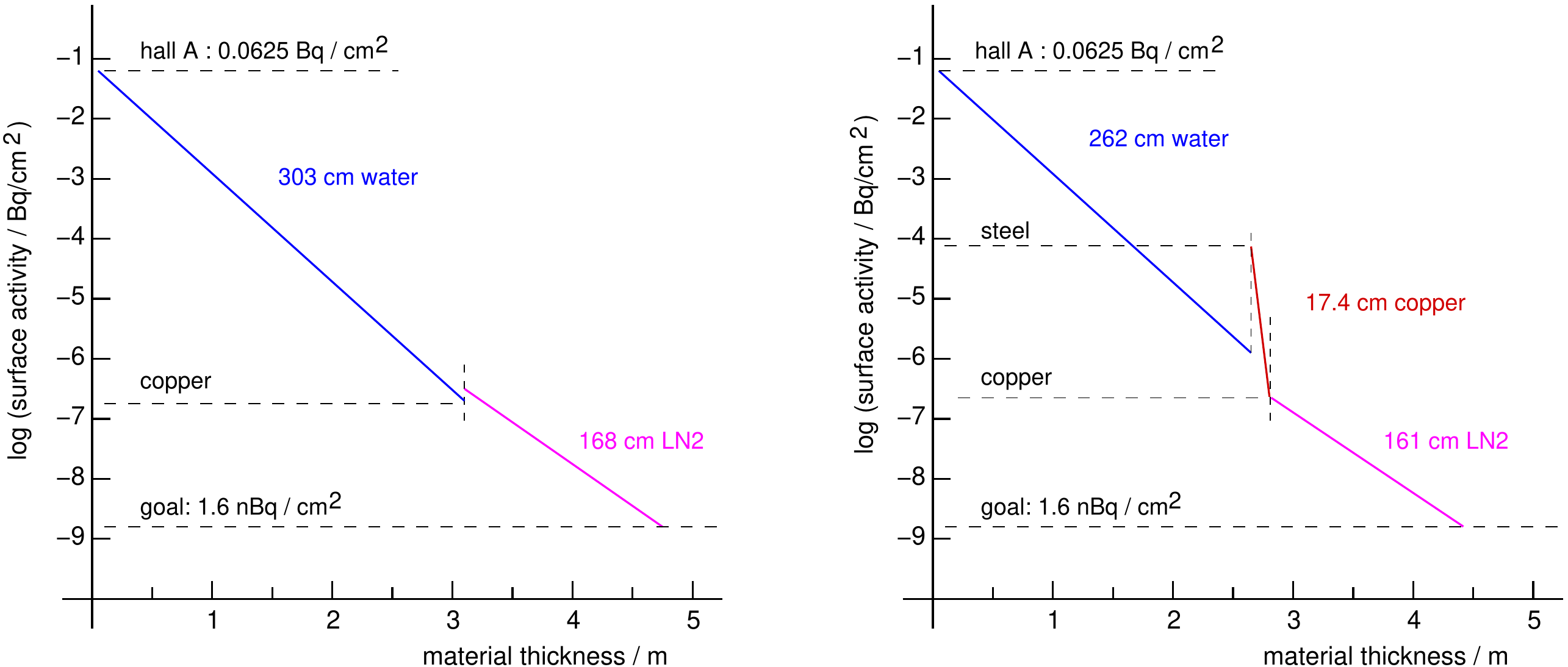}
\caption{
Surface activities as a function of shielding thickness for the graded shields of the \GERDA\ 
baseline design of a copper cryostat (left) and the
alternative of a stainless steel cryostat with an internal copper shield (right). 
Thicknesses of water, copper and \ln \ layers are indicated.
}
\label{fig:gradedshields}
\end{center}
\end{figure}

Figure~\ref{fig:gradedshields} shows on the left a graded shield of water, copper of $t=3$~cm thickness
as construction material for the cryostat, and \ln .
The thickness of the water layer is chosen such to attenuate the external 2.615 MeV \gam \ flux to
the level of the copper surface activity. To reduce the resulting low surface activity of 
0.3\,$\mu$Bq/cm$^2$ to the desired surface activity of 1.6\,nBq/cm$^2$, the \ln \ layer has
to be 168\,cm thick. Taking the radial extension of the Ge detector array into account, the inner 
diameter of the cryostat has to be 3.9\,m to achieve the desired BI. 
Preparations for the construction 
of an electron-beam welded superinsulated cryostat of this size from low-radioactivity OFE copper 
(\th228 \ activity below 20\,$\mu$Bq/kg) were well in progress when an unexpected increase of cost and safety 
concerns stopped the project.

%%%%%%%%%%%%%%%%%%%%%%%%%%%%%%%%%%%%%%%%%%%%%%%%%%%%%%%%%%%%%%%%%%%%%%%%%%%%%%%%%%%%%
%%%%%%%%%%%%%%%%%%%%%%%%%%%%%%%%%%%%%%%%%%%%%%%%%%%%%%%%%%%%%%%%%%%%%%%%%%%%%%%%%%%%%%%%%%%%%%%%

The  alternative design of a cryostat made from stainless steel (\th228 \ activity of 20 mBq/kg, typical
for steel used in the Borexino experiment)
would require a \ln \ layer of 351\,cm thickness that, together with the water layer, would not 
fit into \hallA . Even the equivalent thickness of an LAr layer of 217\,cm would barely be acceptable
and would result in a total diameter that prohibits road transport.
An additional copper layer of 17.4\,cm thickness will, however, reduce the surface activity of the 
steel to 0.24~$\mu$Bq/cm$^2$ so that a \ln \ layer of 161\,cm thickness is needed 
(Fig.~\ref{fig:gradedshields}, r.h.s.). 
The thickness of the water shield has to be chosen such that it suppresses the external \th228 \ 
activity to a fraction of the steel surface activity. For a 1\%  fraction the required water thickness 
is 262\,cm. In this case, the design of the cryostat has to accomodate also the additional weight of 
the internal copper shield which is about 50 tons. 
%The replacement of \ln \ by LAr would reduce the BI to about 2$\cdot$\vctsper . 

The above estimates have been refined by detailed Geant4 simulations with a cryostat geometry very close 
to that actually built \cite{igorB}.  The neck through which the detector array is deployed has no water 
shielding which is ignored in above estimates. The full simulation showed however that the $\gamma$ flux 
through the neck is dominating the background for \ln\ filling. This result together with the advantage
of a reduced Cu thickness for LAr 
and the additional background suppression
by the detection of argon scintillation light led to the decision to drop the \ln\ option.
The final solution is a double-walled superinsulated stainless steel cryostat of cylindrical 
shape with an internal copper shield and a 172\,cm long neck (see next section). With  a LAr fill the  
surrounding \th228 \ radioactivity is expected to contribute $8\cdot10^{-6}$\,\ctsper \ 
to the BI \cite{igorB}.

\subsection{Safety considerations}

The storage of a large amount of cryoliquid underground implies the risk to create large volumes of N$_2$ or 
Ar gas in case of leakage which themselves are not poisonous but might cause suffocation
of people working in the underground area. The standard mitigation of this risk is monitoring the
O$_2$ content in air, and increased ventilation in case of too low O$_2$ concentration. The operation
of a cryostat in water constitutes % a significantly 
an enhanced risk since leakage would imply here
both the possibility of a fast evaporation of the cryoliquid or the mixing of water and \LN2 /LAr
resulting in an even faster possibly explosive phase transition. Hence it is obvious that the risk of
leakage should be minimized, as much as possible, by design and construction. These measures are summarized
below while the mitigation of leakage and the safety aspects of the whole system are discussed 
in section \ref{sec:safetyaspects}.

The design and production of the cryostat have been done according to the European Pressure Equipment Directive 
(PED 97/23/EC) for a nominal overpressure of 1.5$\cdot$10$^5$\,Pa, even though it is operated at 
0.3$\cdot$10$^5$\,Pa and hence below the limit of 0.5$\cdot$10$^5$\,Pa 
above which this code applies. An additional safety margin is achieved by the increase of 3\,mm of 
the wall thickness of the inner (cold) vessel which allows its evacuation. 
Evacuation is needed for He leak tests, Rn emanation measurements and, before the filling with LAr, the
removal of contaminations.
No ports below the filling levels 
of cryoliquid and water 
have been implemented. Standard cryostats have ports e.g. at the bottom for emptying or for
measuring the hydrostatic pressure which are known 
\cite{wit88,ait88} 	%\cite{bar85} 
to increase the risk of leakage.
The entire cryostat is rotational symmetric so that local stress is minimized. Where this symmetry is broken
locally, i.e. at the location of the support and centering pads, the wall thickness is enlarged to
lower absolute stress values. Different from standard cryostats where the inner cold vessel is hung at the 
neck, it is resting here on pads at the bottom. This reduces stress peaks from loads due to earthquakes.
In fact, the cryostat is designed to withstand loads from the \lq Maximum Credible Earthquake\rq \ \cite{sguide} 
of up to 0.6\,g horizontally and vertically without damage. 
The use of highly ductile stainless steel as exclusive construction material enforces the 
\lq leak-before-break\rq \ concept \cite{LBB}. The choice of a stainless steel alloy with Mo and Ti additions, 
1.4571 (X6CrNiMoTi17-12-2), warrants improved corrosion resistance. The
inspections or certifications specified by the PED guarantee a high production
quality that is enhanced by the special requirement of 100\% X-ray inspection of the accessible welds. 

%%%%%%%%%%%%%%%%%%%%%%%%%%%%%%%%%%%%%%%%%%%%%%%%%%%%%%%%%%%%%%%%%%%%%%%%%%%%%%%%%%%%%%%%%%%%%%%%%

%%%%%%%%%%%%%%%%%%%%%%%%%%%%%%%%%%%%%%%%%%%%%%%%%%%%%%%%%%%%%%%%%%%%%%%%%%%%%%%%%%%%%%%%%%%%%%%%%

\section{Engineering description of cryostat}
\subsection{Layout}
\begin{figure} [h]
\begin{center}
\includegraphics[width=0.95\textwidth] {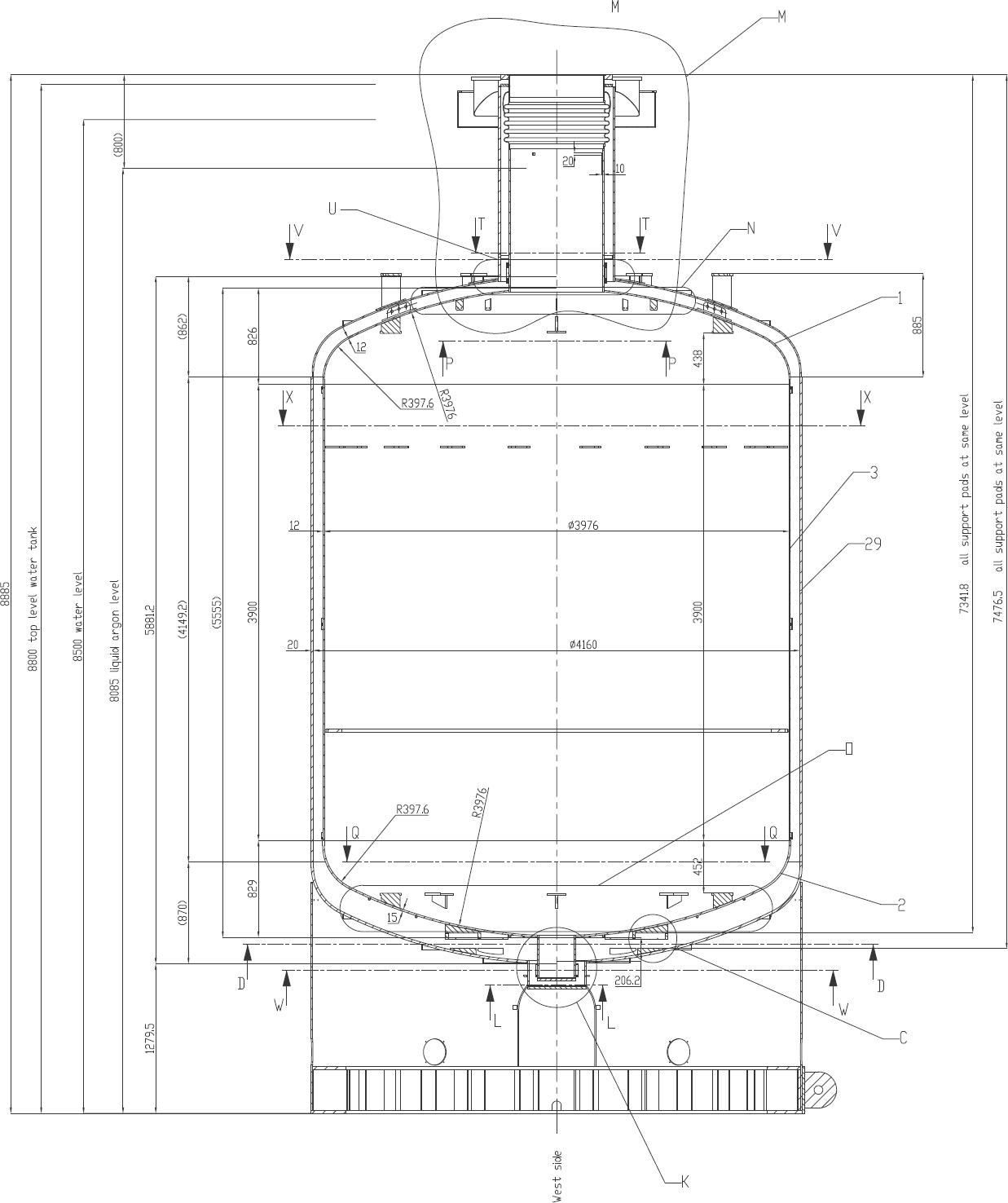}
\caption{Cross section of the \GERDA\ cryostat. Labelled arrows and lines refer to
(not shown) further sectional views and to the parts list, respectively.    
}
\label{fig:cryostat-xsection}
\end{center}
\end{figure}

\begin{figure} [h]
\begin{center}
\includegraphics[width=0.9\textwidth] {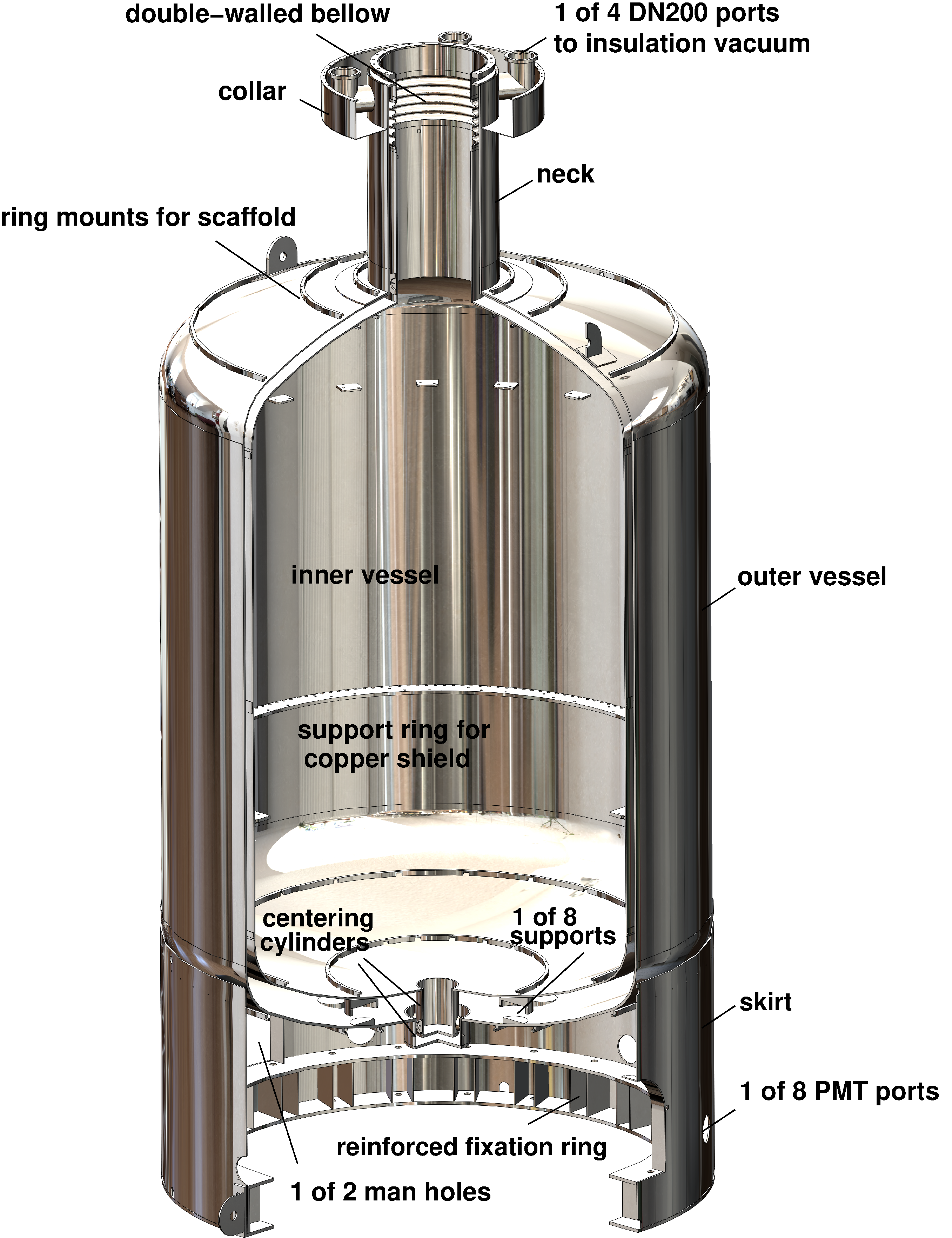}
\caption{CAD drawing of the steel parts of the \GERDA\ cryostat.
}
\label{fig:cryostat-solid-cut}
\end{center}
\end{figure}

Views of the cross section and a solid model of the cryostat\footnote{An alternative layout based on an available standard cryostat 
design has been provided by PJSC \lq\lq Cryogenmash\rq\rq \ 143907, Balashikha, Moscow Region, Russia.}
are shown in Fig.~\ref{fig:cryostat-xsection} and Fig.~\ref{fig:cryostat-solid-cut};
its main characteristics are listed in Table~\ref{tab:cryostatchar}. 
%%%%%%%%%%%%%%%%%%%%%%%%%%%%%%%%%%%%%%%%%%%%%%%%%%%%%%%%%%%%%
\begin{table}[htb]
\begin{center}
\caption{Characteristics of cryostat.}
\vskip+2truemm
\label{tab:cryostatchar}
\begin{tabular}{lcr}
\hline
\hline \\[-2.0ex]
{\bf Materials, Th-228 radiopurity}  \\[0.5ex]
\hline \\[-2.0ex]
vessels, compensators   & 1.4571* (\,25 ton\,)        & 0.1-5~mBq/kg          \\
multilayer insulation  & alu. polyester (15\,kg)      & $<$10~mBq/kg \\
support/centering pads & Torlon\REG \  (8 $\times$ 2.6\,kg ) & $<$5~mBq/kg \\
Belleville springs     & Inconel X718 (8 $\times$ 4.2\,kg)   & not avail. \\
internal shield        & OFRP copper (16 ton)      & 20~$\mu$Bq/kg \\[1ex]
{\bf Geometry}  \\
\hline \\[-2.0ex]
overall height $\times$ diameter    & 8.89 $\times$ 4.20 & [m]     \\
neck height $\times$ inner diameter & 1.72 $\times$ 0.80 & [m]     \\
nominal volume                      & 64                 & [m$^3$] \\
wall thickness inner vessel         & 12, 15             & [mm]    \\
wall thickness outer vessel         & 12, 18, 20         & [mm]    \\
volume bet. inner \& outer vessel  & 8                  & [m$^3$] \\
LAr fill level                      & 6.81               & [m]     \\[1ex]
{\bf Masses} \\
\hline \\[-2.0ex]
empty vessel                        & ~25                & [ton] \\
max. load inner vessel LAr/Cu       & 90~/~48            & [ton] \\[1ex]
{\bf Pressures} \\
\hline \\[-2.0ex]
inner vessel max./min. pressure**          & 1.5~/~-1.          & [10$^5$Pa] \\
outer vessel max./min. pressure**          & 1.0~/~-1.78        & [10$^5$Pa] \\[1ex]
{\bf Multilayer insulation} \\
\hline \\[-2.0ex]
number of layers                    & 30                 & \\
thermal loss                        & $<$300             & [W] \\
active cooling power                & ~500               & [W] \\[1ex]
{\bf Construction code}             & AD2000, DGRL 97/23/EC , & cat IV, mod G \\
\hline \\[-2.0ex]
fraction of x-rayed welds           & 100\%, final orb. weld except                 \\[1ex]
{\bf Earth-quake tolerance}         & 0.6g hor.~\&~vert.      \\
\hline
\hline\\[-1.5ex]
*  X6CrNiMoTi17-12-2  \hfill \\
** relative to atmospheric pressure \hfill 
\end{tabular}
\end{center}
\end{table}
%%%%%%%%%%%%%%%%%%%%%%%%%%%%%%%%%%%%%%%%%%%%%%%%%%%%%%%%%%%%%%%%%%%%%%%%%%%%%%

%%%%%%%%%%%%%%%%%%%%%%%%%%%%%%%%%%%%%%%%%%%%%%%%%%%%%%%%%%%%%%%%%%%%%%%%%%%%%%%%%%%%%%%%%%%%%%%%
%%%%%%%%%%%%%%%%%%%%%%%%%%%%%%%%%%%%%%%%%%%%%%%%%%%%%%%%%%%%%%%%%%%%%%%%%%%%%%%%%%%%%%%%%%%%%%%%

%\clearpage

%%%%%%%%%%%%%%%%%%%%%%%%%%%%%%%%%%%%%%%%%%%%%%%%%%%%%%%%%%%%%%%%%%%%%%%%%%%%%%%%%%%%%%%%%%%%%%%%
%%%%%%%%%%%%%%%%%%%%%%%%%%%%%%%%%%%%%%%%%%%%%%%%%%%%%%%%%%%%%%%%%%%%%%%%%%%%%%%%%%%%%%%%%%%%%%%%

The cryostat\footnote{Fabricated by Simic S.p.A., I-12072 Camerana, Italy.}
consists of two coaxial vessels that are built from torospherical 
heads\footnote{Produced by ANTONIUS, NL-6051 Al Maasbracht, The Netherlands.} 
of 3976\,mm and
4160\,mm inner diameter and corresponding cylindrical shells of 3900\,mm and 4149\,mm height. Each vessel
has a cylindrical neck of 789\,mm and 964\,mm inner diameter, respectively, and of about 1.7\,m height;
on top they are connected to each other by a 
%massive 
flange which will hold a full metal seal.
Since the inner container rests via pads on the bottom of the outer vessel (see below), its shrinkage of
about 2\,cm when cooled down from 300\,K to 77(88)\,K is compensated by a double-walled stainless steel
expansion joint\footnote{Delivered by HKS Unternehmensgruppe, D-18057 Rostock, Germany.} 
in its neck. It is kept centered by six Torlon\REG \ spacers within the neck and at the
bottom. 
In case of earthquake the largest stress will occur at the transition between neck and upper vessel head,
and thus this part is reinforced by an additional stiffening ring.

The volume between inner and outer vessel, about 8\,m$^3$, will hold the vacuum multilayer insulation.
Access to this volume is provided by four CF200 flanges at the very top of the neck which are used
as connections to the pumping station, safety and pressure relieve valves and manometers
(see Fig.~\ref{fig:cryostat-solid-cut}).
The flanges are separated from the water tank by a circular collar around the neck that represents also the
support ring for the flexible fabric which closes the gap between cryostat and water tank.

The inner (cold) vessel is designed for vacuum and 1.5$\cdot10^5$\,Pa overpressure (i.e. 
2.5$\cdot10^5$\,Pa absolute pressure) together with the hydrostatic pressure of LAr. 
The outer vessel is designed for 
1.0$\cdot10^5$\,Pa overpressure and vacuum together with an external pressure of a water column of
7.8\,m (-1.78$\cdot10^5$\,Pa external overpressure). The construction material, type 1.4571 stainless 
steel, 
%(X6CrNiMoTi17-12-2)
purchased in various batches of varying thickness for vessel heads and cylindrical walls, has been
screened  by  \gam \ spectroscopy \cite{maneschg}, and the results for the crucial \th228 \ activity are
shown in Table~\ref{tab:steelscreen}.

A cylindrical skirt made from 12\,mm thick stainless steel holds the cryostat 1280\,mm above floor.
Two manholes provide access to the volume underneath the cryostat, and 8 circular ports allow this volume to
be watched by photomultipliers of the muon veto system \cite{muonveto}.
The bottom of the skirt is reinforced with an annular
structure by which the cryostat is attached to the floor with 24 M39 stainless steel bolts arranged
equidistantly on a bolt circle of 4\,m diameter.

%\clearpage

%\clearpage

%%%%%%%%%%%%%%%%%%%%%%%%%%%%%%%%%%%%%%%%%%%%%%%%%%%%%%%%%%%%%%%%%%%%%%%%%%%%%%%%%%%%%%%%%%
\begin{table}[htb]
\begin{center}
\caption{Size and activity (in mBq/kg) of 1.4571 stainless steel sheet material used for vessel production.
Full screening information is provided for each sample in \cite{maneschg}.  
}
\vskip+2truemm
\label{tab:steelscreen}
\begin{tabular}{lcrcr}
\hline
\hline \\[-2.0ex]
{\bf Outer vessel} & Sample & L x H x t [mm$^3$] & \th228 &  \co60 \\[0.5ex]
\hline \\[-2.0ex]
top head   &D3  &4800 x 2500 x 12 & $1~\pm$~0.4  & 15 \\
           &D3  &4800 x 2500 x 12 & $1~\pm$~0.4  & 15 \\	   
wall       &G5  &6000 x 2000 x 20 & $1.5~\pm$~0.2& 16 \\
           &G1  &1900 x 2000 x 20 & $<0.2$      & 46 \\
           &D6  &6000 x 2000 x 20 & $<0.8$      & 17 \\
	   &G1  &6000 x 2500 x 20 & $<0.2$      & 46 \\
	   &G2  &8000 x 2500 x 20 & $<0.1$      & 14 \\
bottom head&D1  &4800 x 2500 x 20 & $3.4~\pm$~1    & 7 \\ 
	   &D1  &4800 x 2500 x 20 & $3.4~\pm$~1    & 7 \\[0.5ex]
{\bf Inner vessel} \\
\hline \\[-2.0ex]
top head   &D2  &4600 x 2500 x 12 & $<1.7$      & 14 \\
           &D2  &4600 x 2500 x 12 & $<1.7$      & 14 \\
wall       &G3 &12900 x 2000 x 12 & $<0.4$       & 14 \\
	   &G3 &12900 x 2000 x 12 & $<0.4$       & 14 \\
bottom head&D5  &4600 x 2500 x 15 & $<1.1$       & 17 \\ 
	   &D4  &4600 x 2500 x 15 & $<1.8$       & 15 \\
\hline
\hline
\end{tabular}
\end{center}
\end{table}
%%%%%%%%%%%%%%%%%%%%%%%%%%%%%%%%%%%%%%%%%%%%%%%%%%%%%%%%%%%%%%%%%%%%%%%%%%%%%%%%%%%%%%%%%%%%%%%%%%%

\subsection{Support of inner vessel}
The inner vessel rests on 8 support pads. The respective
horizontal counter adaptors are welded equally spaced at the bottom
vessel heads of inner and outer vessels on a circle of 1600~mm diameter and milled
afterwards for good flatness. The inner vessel is kept centered by six Torlon\REG \ spacers within 
the neck and at the bottom. 
%%%%%%%%%%%%%%%%%%%%%%%%%%%%%%%%%%%%%%%%%%%%%%%%%%%%%%%%%%%%%%%%%%%%%%%%%%%%%%%%%%%%%%%%%%%%%%%%
\begin{figure} [h]
\begin{center}
\includegraphics[width=.6\textwidth] {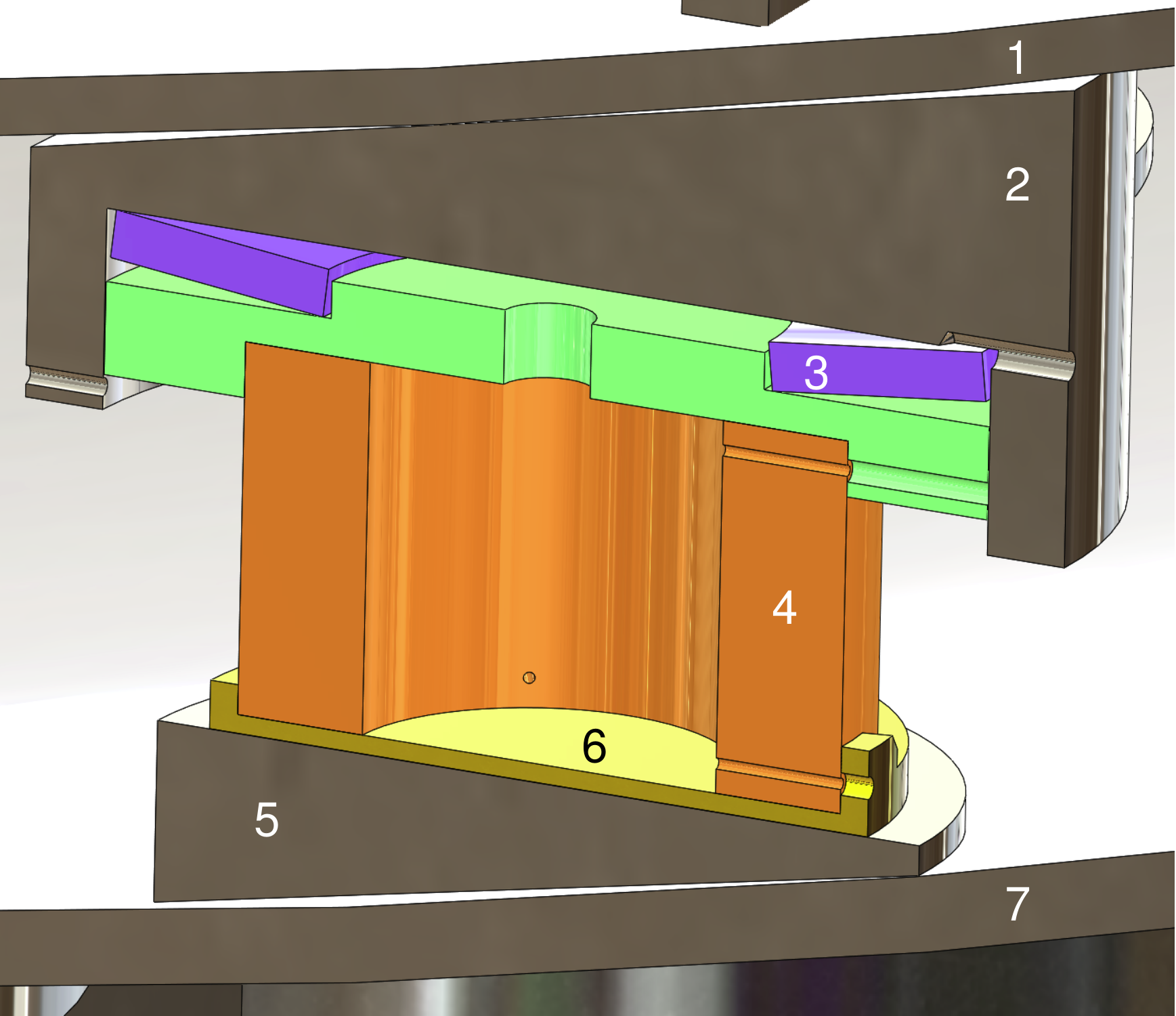}
\caption{Close up of one of the eight supports on the bottom of the inner vessel consisting of the cage (2) for
the Belleville spring (3) welded to the bottom vessel head of the inner vessel (1), the Torlon\REG \
cylinder (4) for thermal insulation, and the countersurface (5) at
the bottom vessel head (7) of the outer vessel.
The bronze shoe (6) for improved sliding has not been implemented.
}
\label{fig:belleville}
\end{center}
\end{figure}
%%%%%%%%%%%%%%%%%%%%%%%%%%%%%%%%%%%%%%%%%%%%%%%%%%%%%%%%%%%%%%%%%%%%%%%%%%%%%%%%%%%%%%%%%%%%%%%%
Fig.~\ref{fig:belleville} shows a schematic of one support pad. 
The usual material
for support pads, glass fiber reinforced epoxy, has been replaced by 
%cylinders)
tubes (\O 183(100) $\times$ H~100 mm) made from Torlon\REG \ 4203/4503. At 77\,(296)\,K this  polyamide-imide 
(PAI)
material exhibits a better radiopurity of $<$14\,mBq/kg (\th228 ), a lower thermal conductivity of
0.1\,(0.26)~W/(m$\cdot$K \cite{barucci}), and comparable flexural strength of 282 (244) MPa \cite{solvay}). 
Belleville springs\footnote{Delivered by MUBEA, D-57567 Daaden, Germany.} made from Inconel X718 with spring constants of 58~kN/mm 
are used to compensate small differences of the gaps between the bearings so that 
all pads will carry about equal load once the cryostat is filled (see sect. 6.1).  Each pad is 
fixed at the cold inner vessel. During cool down, the warm end of the 
pad will move by about 3 mm.  
%Because of the horizontal adaptors only vertical, compressive loads, will act on the Torlon\REG \ pads.

\subsection{Multilayer superinsulation}

\begin{figure} [htb]
\begin{center}
\includegraphics[width=0.7\textwidth] {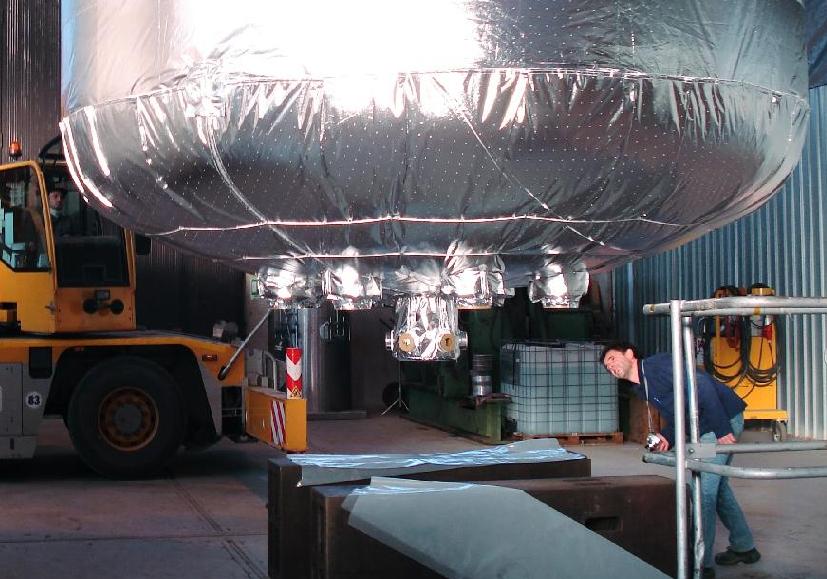}
\caption{Detail of the MLI arrangement at the bottom of the inner vessel showing the
         coverage of support and centering pads.    
}
\label{fig:mli}
\end{center}
\end{figure}

The complete outer surface of the inner vessel is covered up to the compensator with multilayer
superinsulation (MLI) consisting out of 2x 15 
layers.\footnote{Delivered and mounted by Jehier, F-49120 Chemille, France.} 
The nominal thickness of a 15 layers MLI blanket is 3 to 5 mm. The inner and outer layers of a 
blanket are made from TERIL 53, a 6~$\mu$m thick polyester film aluminized on both sides with a 
thickness of 400 \AA, reinforced with a polyester net. The other 13 layers consist of IR 305, i.e. 
a 6 $\mu$m thick perforated polyester film aluminized on both sides with a thickness of
400 \AA, and a polyester tulle with an approximate weight of 5 g/m$^2$. The perforation diameter
is 2~mm, and the distance between perforations is 56 mm. Fig.~\ref{fig:mli} shows the bottom part
of the inner vessel including the Torlon\REG \  pads with the MLI assembly.
%%%%%%%%%%%%%%%%%%%%%%%%%%%%%%%%%%%%%%%%%%%%%%%%%%%%%%%%%%%%%%%%%%%%%%%%%%%%%%%%%%%%%%%%%%%%%%%%
%%%%%%%%%%%%%%%%%%%%%%%%%%%%%%%%%%%%%%%%%%%%%%%%%%%%%%%%%%%%%%%%%%%%%%%%%%%%%%%%%%%%%%%%%%%%%%%%
%%%%%%%%%%%%%%%%%%%%%%%%%%%%%%%%%%%%%%%%%%%%%%%%%%%%%%%%%%%%%%%%%%%%%%%%%%%%%%%%%%%%%%%%%%%%%%%%
Table~\ref{tab:thermloss} shows that the 8 Torlon\REG \  pads and the superinsulation contribute similarly
to the expected thermal loss of about 250\,W while the neck - due to its thin walled (2x 0.8\,mm thick) 
bellow - contributes 10\% only. 
%%%%%%%%%%%%%%%%%%%%%%%%%%%%%%%%%%%%%%%%%%%%%%%%%%%%%%%%%%%%%%%%%%%%%%%%%%%%%%%%%%%%%%%%%%%%%%%%
\begin{table}[h]
\begin{center}
\caption{Contribution of various components to thermal losses. 
}
\vskip+2truemm
\label{tab:thermloss}
\begin{tabular}{lccr}
\hline
\hline \\[-2.0ex]
Component       & area, height         & thermal conductivity & thermal loss \\[0.5ex]
\hline\\[-2.0ex]
neck, bellow    & 0.0041\,m$^2$, 0.44\,m  & 15\,W/(m$\cdot$K)    &   30~W\phantom{*} \\
Torlon\REG \  pads     & 8x0.041\,m$^2$, 0.1\,m  & 0.18\,W/(m$\cdot$K)  & 128~W\phantom{*}   \\
superinsulation &  100 m$^2$, -           & 0.5 W/m$^2$          & 100 W*  \\
\hline
\hline
\end{tabular}
\end{center}
\vskip-2mm
~~~~~~~~~~* including safety factor of 2
\end{table}

%%%%%%%%%%%%%%%%%%%%%%%%%%%%%%%%%%%%%%%%%%%%%%%%%%%%%%%%%%%%%%%%%%%%%%%%%%%%%%%%%%%%%%%%%%%%%%%%

\subsection{Additional thermal insulation}
Inner and outer vessel carry thermal shields which keep the Ar evaporation rate below
10$^4$~m$^3$/h (STP) for the case of a water/LAr leak in the outer/inner vessel 
(see section \ref{sec:safetyaspects}).

The outside surface of the vertical shell of the inner vessel is covered below the MLI with a
6~mm thick polycarbonate (Makrolon\REG ). 
This material has been chosen since its properties are adequate for the use at cryogenic temperatures, and since 
it is rigid and has
a relatively low thermal conduction coefficient of 0.2 W/(m$\cdot$K). Less favourable features are its
relatively large linear thermal expansion coefficient (65$\cdot$10$^{-6}$/K at room temperature) as
well as its water absorption of 0.12\% at 23\,\deg \ and 50\% relative humidity.

Similarly, the outer surface of the vertical shell of the outer vessel as well as its bottom are covered
with 6~mm thick extruded polystyorol foam\footnote{Jackodur\,KF300\,FTD by JACKON 
                                           Insulation GmbH, D-33803 Steinhagen, Germany.} 
which is hermetically covered by a multilayer polyester 
foil\footnote{VM2000 by 3M\textsuperscript{TM} Deutschland GmbH, D-41453 Neuss, Germany.}  
with $>98\,\%$  specular  reflectance
that serves in addition as reflector for the Cherenkov radiation in the water tank.

The thickness of the polycarbonate and the polystyrol layers have been deduced from test measurements
of the heat transfer in the pool boiling regime for the interface liquid nitrogen - stainless steel -
water (see section \ref{sec:safetyaspects}).

\subsection{Production engineering}

The production process was subject to the AD 2000 code and PED 97/23/EG and controlled by representatives of a notified 
body (T\"UV Nord, Germany). 
We summarize here a few non-standard additional features for further improvement of the cryostat's reliability
and performance.

The production of the cryostat and the MLI installation have been carried out in a \lq gray\rq \ area 
which was separated from the standard carbon steel production sites. All welds were done by TIG welding using
exclusively thorium-free TIG-welding electrodes. For the welding filler, material of less than 5~mBq/kg 
\th228 \ radioactivity has been 
selected. Before production, the effectiveness of weld preparation and procedures even in case of thermal shocks
was verified with two 500\,mm long 1.4571 sheets that were welded together and subsequently immersed into liquid 
nitrogen. The chosen welding procedure yielded for the low temperature notch impact energy the value of 120\,J that is 
almost 4 times higher than required by AD2000. All welds have been passivated to avoid corrosion. The cleaning procedure
for all inner and outer surfaces included removal of oil and grease, pickling and passivation, rinsing with de-ionized water
and subsequent drying. Ferroxyl tests did not reveal any ferrite inclusions on the surfaces. 

All accessible welds of  the inner and outer container of the cryostat have been 100\% X-ray tested. Where such a test 
was not
possible like at the final orbital weld an ultrasonic test was done.

\subsection{The internal copper shield}

\begin{figure} [htb]
\begin{center}
\includegraphics[width=0.4\textwidth] {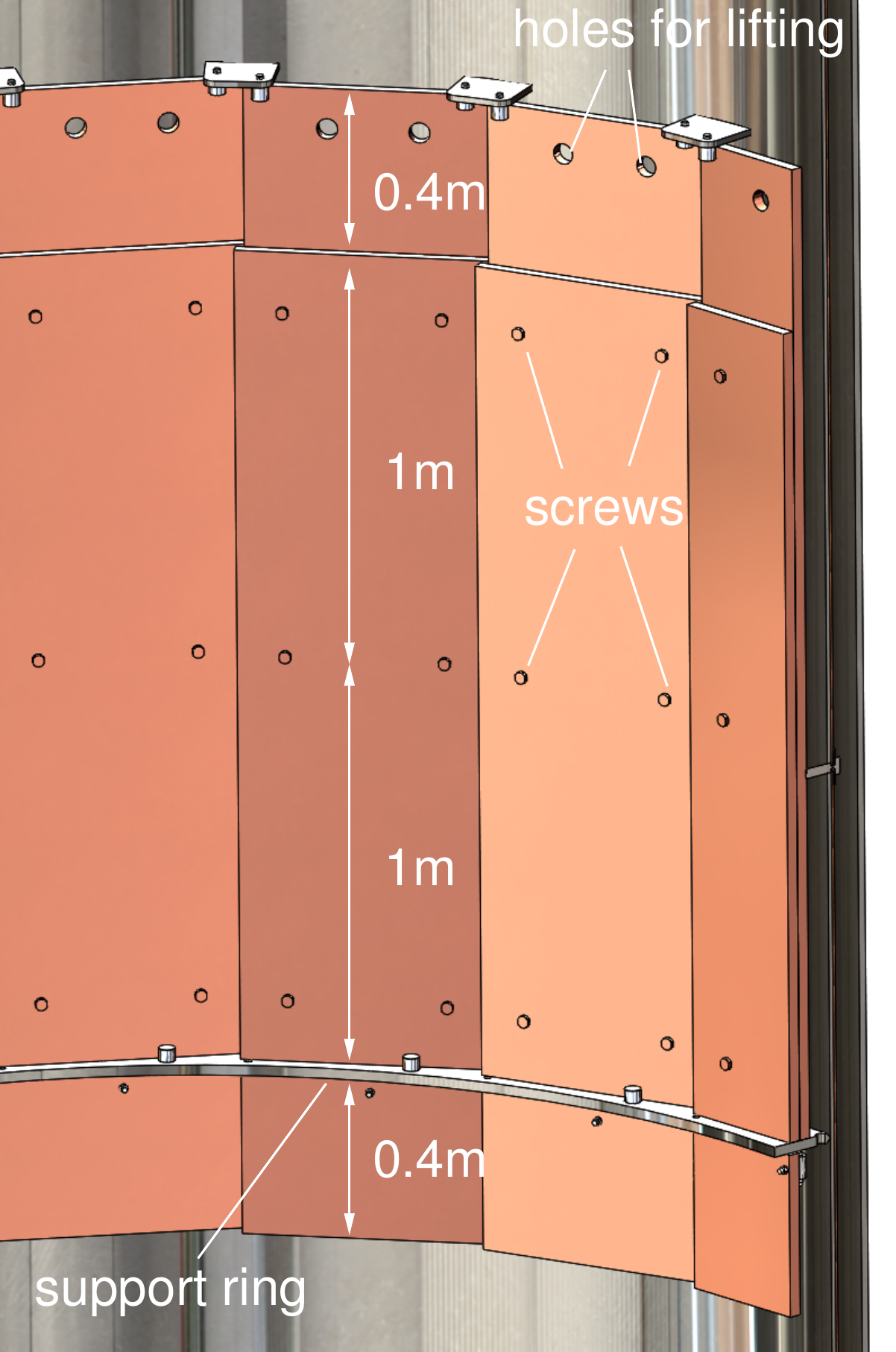}
\caption{Three (and a half) of 20 segments of the internal copper shield as mounted in the cryostat. 
}
\label{fig:Cu-shield1}
\end{center}
\end{figure}

The screening of the stainless steel material used for the cryostat yielded an unexpected low 
radioactivity \cite{maneschg} that
allowed to reduce the amount of the internal copper shield from the envisaged 48 tons to less
than 16 tons. Its thickness and profile have been determined by Monte Carlo calculations \cite{igorB}.
The profile is symmetric w.r.t. the midplane, the thickness is 6 cm up to the height of 1 m,
and continues from there with 3 cm thickness up to 1.4 m (see Fig.~\ref{fig:Cu-shield1}). 
The shield is assembled from 20 overlapping segments which fit through the cryostat's neck. 
Each segment consists of three 3 cm thick and 61.5 cm wide copper plates of 40, 200 and 240\,cm
length, respectively. The two longer plates are screwed together and  rest on a support
ring within the cryostat. The short plate is attached below.
%%%%%%%%%%%%%%%%%%%%%%%%%%%%%%%%%%%%%%%%%%%%%%%%%%%%%%%%%%%%%%%%%%%%%%%%%%%%%%%%%%%%%%%%%%%%%%%%
%%%%%%%%%%%%%%%%%%%%%%%%%%%%%%%%%%%%%%%%%%%%%%%%%%%%%%%%%%%%%%%%%%%%%%%%%%%%%%%%%%%%%%%%%%%%%%%%

The copper plates have been rolled\footnote{By CSN Schreiber, D-57290 Neunkirchen, Germany.} in spring 2007 
from freshly produced OFPR copper.\footnote{By Norddeutsche Affinerie, now Aurubis AG, D-20539 Hamburg, Germany.}  
After rolling, the plates were warmed up to 
50$^\circ$\,C, pickled in $>$15\% sulfuric acid, and, immediately afterwards, rinsed with de-ionized water.
The individual segments were vacuum-packaged after assembly, and an additional plastic wrap protected them for 
transportation to \LNGS \ where they are kept underground since June 2007 to prevent further cosmic activation. 

%\section{Cryogenic infrastructure}

%\input{CC201228-CRYOSYSTEM.tex}

\section{Cryogenic infrastructure}
The cryogenic infrastructure has to ensure a stable and safe
%%%%%%%%%%%%%%%%%%%%%%%%%%%%%%%%%%%%%%%%%%%%%%%%%%%%%%%%%%%%%%%%%%%%%%%%%%%%%%
\begin{figure}
\begin{center}
\includegraphics[width=0.7\textwidth] {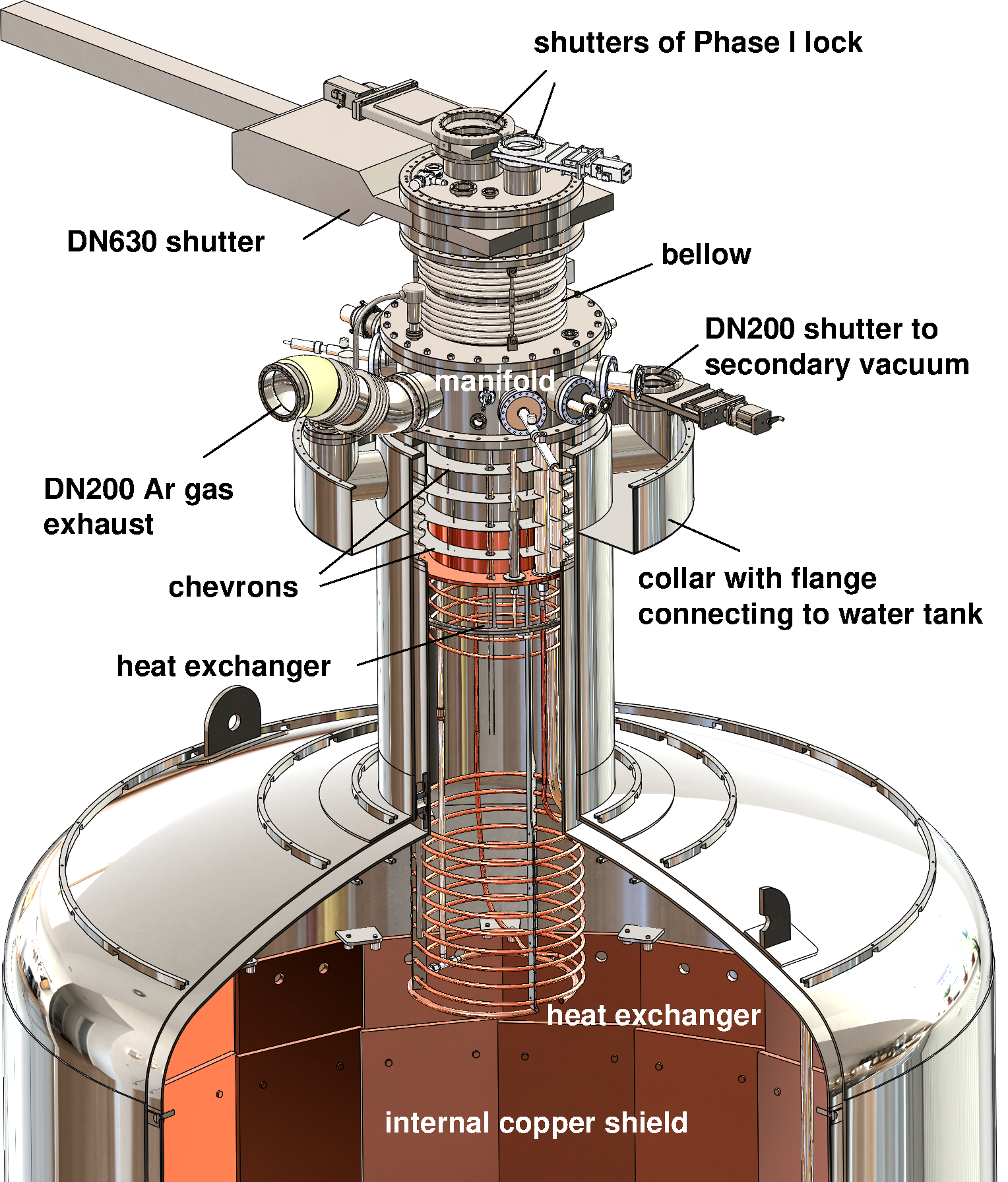}
\caption{Close up of the top of the cryostat including neck, manifold and interface to the lock.
For \GERDA \ Phase II, the two shutters of the Phase I lock above the DN\,630 shutter have been removed.}
\label{fig:closeupneck}
\end{center}
\end{figure}
%%%%%%%%%%%%%%%%%%%%%%%%%%%%%%%%%%%%%%%%%%%%%%%%%%%%%%%%%%%%%%%%%%%%%%%%%%%%%%%%
operation of the cryostat.
Since the neck provides the only access to the interior of the cryostat,
a manifold on top of the neck carries the flanges for all lines that
penetrate into the cold volume including filling tube, gas exhaust tube, tubes for
active cooling, and feedthroughs for the cryostat instrumentation
(see Fig.~\ref{fig:closeupneck}). 
  The manifold is connected with a bellow to a DN630 shutter\footnote{Gate valve, Series 19, 
  by VAT Deutschland, D-01109 Dresden, Germany.} 
  which
  is located at the floor of the clean room. The bellow allows for
  25 mm lateral movements of the clean room with respect to the cryostat in
  case of an earth quake.  The shutter separates the
  cryostat from the lock. It is closed while the lock is open
  for germanium detector installation.
All 
valves\footnote {Cryogenic valves by Flowserve K\"ammer, D-25524 Itzehoe, Germany;
for gases manual valves series BG by Swagelok, Solon, Ohio 44139, USA.} 
have metalic sealing against atmosphere.
The control of most pneumatic valves
and readout of sensors is done with a programmable logic
controller (PLC) Simatic S7\footnote{Simatic S7 by Siemens, D-80333 M\"unchen, Germany.} 
that is dedicated to the control and
monitoring of the cryogenic infrastructure. The power for the entire infrastructure
is connected to an uninterruptible power supply backed by a generator which is either driven 
by normal electric power or a battery. \newline
\noindent
The various parts of the cryogenic infrastructure are discussed below.

\subsection{Cryogenic piping}
 The first filling of the cryostat with LAr and the optional refilling 
 during the standard operation
 is done from a selected storage tank (radon emanation about 5~mBq, operated
 at 3$\cdot10^5$\,Pa)
 which is located in the TIR tunnel, about
 30~m away from the cryostat. At the same location are also 2
 storage tanks for LN2. One (4.0-4.5$\cdot10^5$\,Pa) for LN2 extraction at the nominal pressure
%\footnote{\lq Gauge pressure\rq \ is used to denote the pressure relative to atmospheric pressure.} 
 of 3.2$\cdot10^5$\,Pa relative to atmospheric pressure for the LAr active cooling
 and  one for  gaseous nitrogen at a higher pressure of 8$\cdot10^5$\,Pa which is used for flushing of 
 glove boxes and the operation of pneumatic valves.
For the pipe\footnote{Cryogenic piping, \lq valve box\rq \ and \lq keep cold device\rq \ produced and 
installed by DEMACO, NL-1723 ZG Noord-Scharwoude, The Netherlands.}
 from the storage tank to the cryostat a triaxial pipe 
 was chosen:
 the LAr pipe is the inner pipe (DN\,25) at a LAr boiling temperature of
 $\approx$99 K, the LN2 flows outside the LAr pipe in a DN\,50 pipe
 at a lower temperature of $\approx$91.3 K and the space between the DN\,50 and the
 outer third pipe (DN\,100) holds the vacuum insulation. The lower LN2 temperature allows to subcool
 LAr without freezing it (84 K) and hence to
 reduce argon flash gas losses.

 As a  safety feature the pressure in the LN2 pipe is used to operate
 pneumatic valves at the outlets of the storage tanks. If due to a
 rupture of the pipe, for example, the pressure is too low
 the valves close and the spilling
 of LN2 and/or LAr will be stopped.

 The LAr passes optionally through a charcoal filter to retain radon (about 1 kg mass) 
  which is  cooled with LN2  to about 86 K. Afterwards two PTFE 
  filters\footnote{Fluorogard CTFZ01TPE by Entegris Eastern and Central Europe, 
  D-01109 Dresden, Germany.}
  with 50~nm pore size and 10 inch length are
  added in series to retain any particles.
 
 The filters together with all valves for LN2 and LAr are in a vacuum
 insulated box called \lq valve box\rq .

 Before the LN2 pipe enters the valve box a phase separator called
\lq keep cold device\rq \ allows to remove gaseous nitrogen. \\

\subsection{Active cooling of LAr} 

Inside the cryostat are two LN2/LAr heat exchangers in form of 
spirals made from 18 mm x 1 mm copper pipes: 
one at the liquid/gas surface and one in the main volume of the
cryostat (see Fig.~\ref{fig:closeupneck}). 
The diameter of the spirals is about 760~mm.

The pressure at the gas outlet of
the heat exchanger is about 1.2$\cdot10^5$\,Pa absolute which corresponds to a LN2 
boiling temperature of 79.6~K. The cryostat  pressure is also regulated to 
 1.2$\cdot10^5$\,Pa absolute which corresponds to an LAr boiling temperature
of 89.0~K at the liquid surface. Due to the hydrostatic pressure
the boiling point increases deeper inside the cryostat.
The LAr is cooled to about 88.8~K. 

%%%%%%%%%%%%%%%%%%%%%%%%%%%%%%%%%%%%%%%%%%%%%%%%%%%%%%%%%%%%%%%%%%%%%%%%%%%%%%
\begin{figure}[h]
\begin{center}
\includegraphics[width=0.7\textwidth] {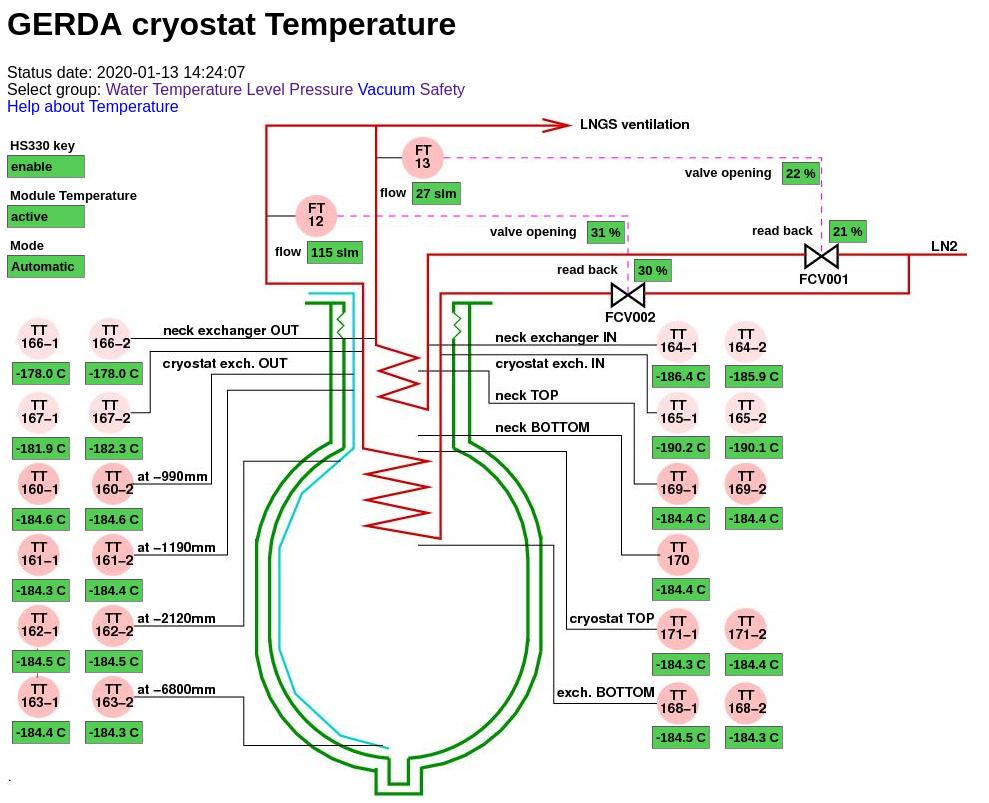}
\caption{
Distribution of temperatures within the LAr cryostat with powered Ge detector
array.
}
\label{fig:cryo-temps}
\end{center}
\end{figure}
%%%%%%%%%%%%%%%%%%%%%%%%%%%%%%%%%%%%%%%%%%%%%%%%%%%%%%%%%%%%%%%%%%%%%%%%%%%%%%%%
There are in total 29 PT-100
sensors inside the cryostat to measure the LN2 temperatures at
the heat exchanger
inlets and outlets as well as the LAr temperatures at different heights.
It turns out (see Fig.~\ref{fig:cryo-temps}) that the LAr temperatures 
are  within the precision of our measurement
(0.1\deg ) the same everywhere
except at the surface where they can be adjusted to be
a few tenths of a degree higher. The interpretation is that
convective currents lead to a strong mixing effect and hence to a homogeneous
temperature. Also: no argon ice formation at the heat exchangers
is observed despite the fact that the LN2 boiling temperature is
about 3 degrees below the freezing point.

The LN2 flow regulating valves are  inside the valve box. The pressure
at the inlet is 3.5-4$\cdot10^5$\,Pa absolute. After the valves the pressure is only
1.2$\cdot10^5$\,Pa absolute. Because of the pressure drop a gas/liquid mixture
passes through the heat exchangers. The diameter of the pipe from
the valve box to the cryostat was chosen to be only 8~mm and the pipe
is always inclining upwards to ensure  a large flow speed such that
the gas/liquid mixture does not separate. This concept \cite{Hstroh} 
is supposed to reduce
flow variations in the cooling and hence microphonics
although the measured gas flow still shows variations
by $\pm 20$\% on a time scale of minutes. Through the
bottom heat exchanger the flow is adjusted to on average 120 slm (liters 
of gas per minute at standard pressure and temperature) and for the other
one to 15 slm. 

%More details of the cryogenic cooling concept are given in \cite{Hstroh}. 
 
\subsection{Vacuum system}

The insulation vacuum is pumped by a 600\,$\ell$/s 
turbomolecular pump\footnote{Turbovac 600C by Oerlikon Leybold Vacuum, D-50968 K\"oln, Germany.} 
with mechanical rotor suspension that was backed at full gas load by a 40\,m$^3$/h , and later by a 8.5\,m$^3$/h 
rotary vane pump. The pump is mounted with a DN160/200CF adaptor onto 
a DN200CF shutter\footnote{UHV gate valve, series 10, by VAT Deutschland, D- 01109 Dresden, Germany.} 
that is attached to one of the four CF200 ports (see Fig.~\ref{fig:closeupneck}).
After some years of operation the turbomolecular pump was replaced by the smaller 150\,$\ell$/s model.   
The turbo pump is running continously. If momentarily switched off, the 
%leakage rate is about
pressure increases at 
5$\cdot$10$^{-7}$\,hPa$\cdot\ell$/sec.

The pressure is monitored redundantly and read out by the PLC.
If it is too high or if, e.g.~the turbo pump signals an error,
the shutter between cryostat and turbo pump will
be closed. The vacuum is also monitored by a residual gas 
analyzer.\footnote{Model PPM100 Partial Pressure Monitor by Stanford Research Systems.} The partial pressures 
for water, argon and nitrogen
are read out and can be used to diagnose a problem in case the
total pressure rises unexpectedly.

Another one of the four CF200 ports carries the overpressure protection device for the insulation
vacuum volume. It is realized by a DN\,100 disk with a rubber O-ring seal which is
tightened by the air pressure. 

\subsection{Pressure regulation}

The Ar gas pressure in the cryostat is one of the critical
parameters. It is regulated by two intelligent pressure transmitters 
that include a PID (proportional, integral, differential) controller
with a mean time between failure (MTBF) of 244 years.\footnote{Model LD301 by Smar Europe B.V., D-55545 Kreuznach, Germany.}
%These intelligent pressure transmitters are quite reliable with a 
%and  have a PID (proportional, integral, differential) controller included. 
Each of the analog output signals is
directly used to control a DN\,50 valve\footnote{Type 2415\,P3 from Flowserve K\"ammer, D-45145 Essen, Germany.} 
($k_{\rm vs}\,=\,40$). These
pressure regulation circuits operate therefore independently of
the PLC which has a lower MTBF rating. 

In case of power failure the above mentioned valves are normally
closed. Therefore a 
DN10 valve\footnote{Type 241\,H3 from Flowserve K\"ammer, D-45145 Essen, Germany.}
with a smaller opening ($k_{\rm vs}\,=\,1.6$) was added.
%which is normally open in case of power failure or loss of 
%nitrogen for the actuator. 
The discharge rate through this `normally open' valve is large enough to
keep the pressure below 1.2$\cdot 10^5$\,Pa absolute for at least one week in
case of loss of power or loss of nitrogen for the actuator. 

\subsection{Safety devices against overpressure}

The cryostat is equipped redundantly with a safety valve and a rupture disk.
The safety valve\footnote{Type 411 from LESER GmbH \& Co. KG, D-20537 Hamburg.} has a 
trigger point 
of 0.85$\cdot 10^5$\,Pa relative to atmospheric pressure and a flow area of 7500 mm$^2$. Since the safety 
valve is not leak tight against the atmosphere,  a DN150
rupture disk\footnote{Type CF160-UKB-LS-FL\,6 from REMBE GmbH Safety + Control, D-59929 Brilon.} 
with a trigger point of 0.8$\cdot 10^5$\,Pa relative to atmospheric pressure has been added in front. 
This disk is welded into its stainless steel holder and has  ConFlat flange DN160 connections.

The same type  is used as rupture disk, here with
a maximum trigger point of 1.4$\cdot 10^5$\,Pa relative to atmospheric pressure. 
To signal its rupture or damage a signal rupture disk is added afterwards. 

Both devices are large enough to cope with a maximum discharge
rate of 4.5 kg/s of 
argon gas. \\

\subsection{Exhaust gas heater}
   The argon exhaust gas is
   passed to the \LNGS \ ventilation system that pushes it through
   a pipe in the highway tunnel to the outside. In case of a faulty operation
   of the cryostat
                            the cold argon gas would freeze the water vapor present
   in the pipe and the fans could stop working. Both
   scenarios would result in a collaps of the system.
   The argon gas must therefore be warmed up above 0$^\circ$\,C
   before it is discharged to the LNGS ventilation system.
A water/gas heat 
exchanger\footnote{Model C300 1708-2PASS by FUNKE W\"armetauscher Apparatebau GmbH, D-31028 Gronau, Germany.}
is installed for this purpose. The exhaust gas from the 
safety valve and rupture disk of the cryostat, from the
safety disk of the insulation vacuum and all other valves 
is collected into  a DN200 pipe which is connected to
the inlet of the heat exchanger. The gas then  passes
through a bundle of 121 pipes with 18 mm inner
diameter. Outside the pipes is
water to heat up the cold gas. The outlet of the
heat exchanger is directly connected to the ventilation
system of \LNGS . A bypass between inlet and outlet is
added with a burst disk to avoid overpressure in case
the heat exchanger is blocked for some reason.

The water is normally the cooling water of \LNGS \ but
as a backup a pump is installed to pump the water from
the water tank through the heat exchanger. 

\subsection{Slow control and graphical user interface} 

%%%%%%%%%%%%%%%%%%%%%%%%%%%%%%%%%%%%%%%%%%%%%%%%%%%%%%%%%%%%%%%%%%%%%%%%%%%%%%
\begin{figure}
\begin{center}
\includegraphics[width=0.8\textwidth] {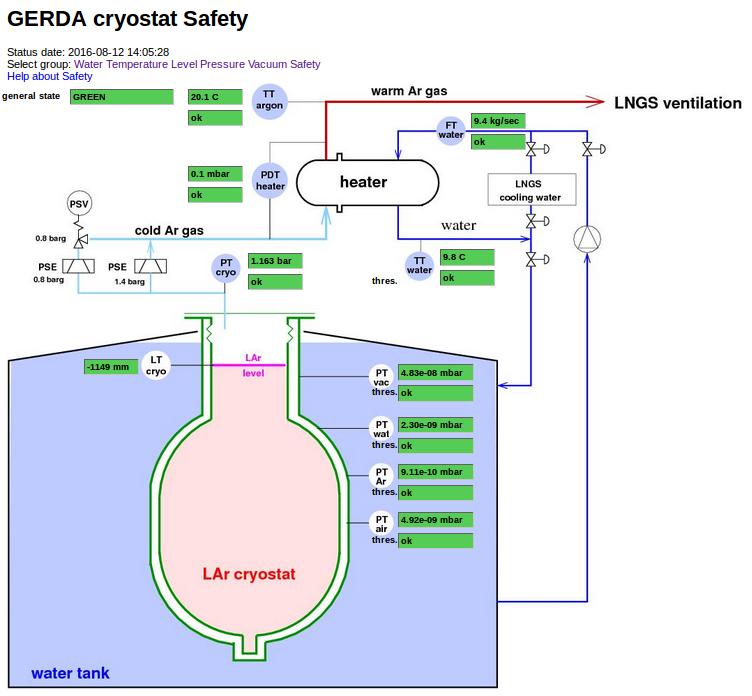}
\caption{Safety status of the GERDA experiment as inquired at 14:05 on August 8, 2016
via the web interface. The values of all critical operation parameters are within
the specifications for standard operation (see subsection~\ref{sec:cops}). }
\label{fig:cryosafety}
\end{center}
\end{figure}
%%%%%%%%%%%%%%%%%%%%%%%%%%%%%%%%%%%%%%%%%%%%%%%%%%%%%%%%%%%%%%%%%%%%%%%%%%%%%%%%
Apart from the redundant high-reliability pressure regulation, all sensors, valves
and pumps of the cryogenic infrastructure are monitored and controlled continously by the 
Simatic S7 PLC.
The PLC evaluates the safety status of the cryogenic system, and automatically triggers 
appropriate actions if needed. The safety relevant status 
information is shared with the \LNGS \ safety network via MODBUS and hardwired by a cable 
transmitting the relevant status bits. Using CGI scripts, a Linux web server reads or writes data 
from or to the PLC via TCP/IP access, acting both as DBclient for the general 
\GERDA \ slow control system \cite{SlowControl} and the Hypertext Transfer Protocol (HTTP). Thus 
performance and status of the \GERDA \ cryogenic system can be examined and, in part, controlled 
remotely. Fig.~\ref{fig:cryosafety} shows 
for example the safety summary page where all critical operation parameters for the cryostat 
and supporting infrastructure are displayed and color-coded (see  subsection~\ref{sec:cops}). 
Other pages show the status of the water system, 
the temperature distribution (see Fig.~\ref{fig:cryo-temps}), the fill level, and the values of the pressures measured 
for cryostat and insulation vacuum.  

\subsection{Water drainage}
 The fast drainage of the water is important in case
 of an emergency. Therefore the PLC that monitors the cryostat is
 also controlling the drainage valves of the water tank. There are
 three different lines:
(i) the main DN250 pipe below the tunnel road with an allowed maximum flow rate of
80\,\lps , (ii) a DN90 line (~18\,\lps ) to 100\,m$^3$ storage containers  
in \hallA , the so-called \lq GNO pits', and (iii) a DN80 pipe (20\,\lps ) which is 
part of the emergency water supply system for the \GERDA\ exhaust heater leading 
also to the GNO pits. The PLC controls the valves to these lines as well as the 
flow rate in the main DN250 pipe via a butterfly valve so that a constant standard
flow rate of 20\,\lps \ or, in emergency, a maxium flow rate of $\leq$80\,\lps \ are 
maintained independent of the hydrostatic pressure in the water tank.

A test of the fully automated drainage procedure in 2010 showed  that measured and 
predicted drainage times are in very good agreement with each other, and that in 
case of emergency the water tank can be emptied in less than 2 hours.

\section{Specific safety aspects}
\label{sec:safetyaspects}

\subsection{Evaporation rates in case of failures}

Compared to  standard cryogenic installations the \GERDA \ cryostat-water-tank system exhibits the
additional risk that the separation between water and the cryogenic container is broken resulting in
potentially huge gas exhaust rates.  

In the worst case scenario, i.e. the simultaneous rupture of inner and outer container wall, the mixing of LAr and water 
could lead to a rapid phase transition. External  
%The risk from a mixing of LAr and water that could lead to a rapid phase transition is evaluated by external
consultants\footnote{Nier Ingegneria S.p.A., I-40013 Castel Maggiore (Bo), Italy.} 
have evaluated this risk and found it to be extremely 
unlikely, $<10^{-8}$\,ev/yr, due to 
(i) the mutual independence of the two stainless steel cryogenic vessels that are built according to 
established construction codes,
(ii) the irrelevance of corrosion related to the galvanic interaction between copper shield and the 
 austenitic 1.4571 stainless steel, 
(iii) the earthquake tolerance of 0.6\,g, 
(iv) the redundant monitoring of the critical operational parameters (COPs, see below sect.~\ref{sec:cops}),  
and last but not least,  
(v) the fulfillment of the \lq leak-before-break' principle. 
 
In case of a small leak in one of the two vessels the insulation vacuum would be lost.
A compilation of the thermal conductivity of superinsulation in dependence of the gas pressure
shows \cite{Tim89} that in the most conservative estimate a superinsulation with a heat flux of 1\,W/m$^2$
between 77 and 300\,K is degraded by a factor of 1000 in case of a complete loss of insulation vacuum.
Hence, with the cryostat's  surface of about 100\,m$^2$ and 97 tons of LAr it will take about 44 hours
until all LAr has been evaporated. The resulting gas flow (20\,C$^\circ$) of about 1340\,m$^3$/h 
is small compared to the limit of 10000\,m$^3$/h set by \LNGS .

A major leak in the outer vessels would destroy not only the insulation vacuum but enable the surrounding 
water to enter the space between inner and outer vessel. Water with its high specific heat capacity
of 4.19\,kJ/(kg$\cdot$K) represents an efficient heater for the cryogenic liquid. A drop of less than 
6$^\circ$\,C in temperature of the 590\,m$^3$ stored water is enough to evaporate the %70
64\,m$^3$ of
LAr contained in the cryostat. An estimate of the resulting gas flow is difficult lacking detailed
knowledge of the involved heat transfer. The heat transfer in an unsteady state involving phase changes
has been discussed by various authors, see e.g. \cite{Van63, Fas70} but the application of the 
reported results is not obvious. Hence, experimental model studies have been performed in order to arrive 
at a more quantitative understanding of the evaporation rates in a worst case scenario for a water-cryostat 
system (see Appendix A). 

%%%%%%%%%%%%%%%%%%%%%%%%%%%%%%%%%%%%%%%%%%%%%%%%%%%%%%%%%%%%%%%%%%%%%%%%%%%%%%%%%%%%%%%%%%%%%%%%
\begin{figure}
\begin{center}
\includegraphics[width=0.8\textwidth] {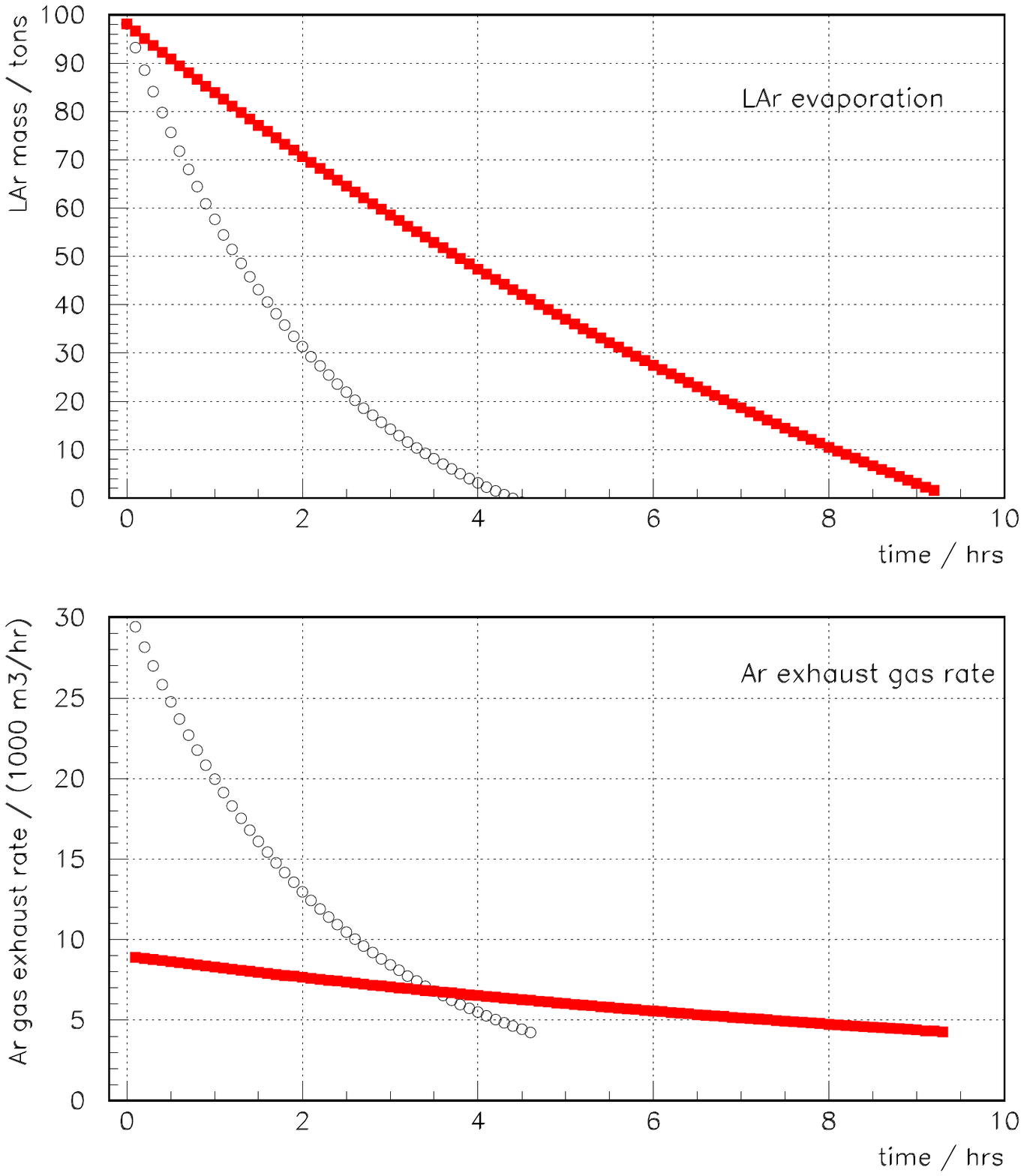}
\caption{
Results of model calculations for the argon mass loss (top) and evaporation rate (bottom) in case
of an emergency with a heat transfer coefficient of 27~kW/m$^2$ (black open circles)
and a reduction to 5 kW/m$^2$ in the cyclindrical part (red filled circles).
}
\label{fig:evasimu}
\end{center}
\end{figure}
%%%%%%%%%%%%%%%%%%%%%%%%%%%%%%%%%%%%%%%%%%%%%%%%%%%%%%%%%%%%%%%%%%%%%%%%%%%%%%%%%%%%%%%%%%%%%%%%
Using the deduced maximum heat transfer coefficient of 27\,kW/m$^2$ for LAr, 
%and the formalism for analyzing the data of the small scale model 
the rate of exhaust gas for the
\GERDA \ configuration with a total mass of 90 tons of LAr is predicted for the worst case scenario, that
is the sudden removal of the outer container such that the inner cryogenic container is immediately 
exposed to the surrounding water. The black circles in Fig.~\ref{fig:evasimu} show that at the assumed heat
transfer coefficient of  27\,kW/m$^2$ the total mass is evaporated after less than 5 h, and that the
corresponding gas exhaust rate peaks at almost 30000\,m$^3$/h which is not acceptable.
However, if the heat transfer through
the cylindrical wall is limited to 5\,kW/m$^2$ (red curves in Fig.~\ref{fig:evasimu}) it will take about 
9 hours to evaporate all LAr,
and the gas exhaust rate will stay well below the limit of 10000\,m$^3$/hr. The limitation of heat transfer
can be achieved by covering the wall of the inner vessel with about 5 to 10\,mm 
thick plastic material of
thermal conductivity
between 0.1 and 0.25\,W/(m$\cdot$K). This requirement is well fulfilled by polycarbonat which exhibits
also the required vacuum compatibility and low temperature properties.  
The amount of LAr contained in the volume above the covered cylindrical wall in the vessel head and neck is
about 3\,m$^3$. Its evaporation would take less than 5 minutes in which time the exhaust rate could 
reach 13000\,m$^3$/h.  

For a leak in the inner vessel similar considerations apply.
The maximum gas flow might be lower since overpressure inside the insulation vacuum volume
might inhibit further flow of the 
cryoliquid. However, other than in case of a leak in the outer shell, the prevailing amount of the gas 
cannot escape through the cryostat's neck but has to pass through the DN100 overpressure
release disk mounted on one of the CF200 ports.  
Thus the insulation volume is also equipped with an overpressure safety device.
The thermal shield does not need to be vacuum compatible. Hence on the outside of the outer vessel
lightweighted 6\,mm thick extruded polystyrol (Jackodur) panels 
(and a VM2000 foil layer for the muon veto) are used. 
 
\subsection{Critical operational parameters, alarms and mitigating actions}
\label{sec:cops}
Table~\ref{tab:COPs} provides a list of the three COPs identified in
the risk analysis \cite{nierrisk}, (i) pressure in the cryogenic volume of the cryostat, (ii) pressure in the insulation volume, and
(iii) the LAr fill level\footnote{This parameter turned out to be irrelevant since an insignificant amount of LAr was lost only during Ge 
detector immersion.}  
in the cryostat's neck. Seven additional COPs  allow superior diagnostics of 
possible failures and the suppression of false alarms: (iv-vi) partial pressures of 'air', argon and water in the
insulation vacuum, and four parameters of the water-cooled Ar gas heater, (vii) differential pressure between
entrance and exit for the Ar gas being directly proportional to the square of the mass flow through the heater,
(viii) temperature of the exhaust gas at its exit, (ix) flow rate of water through heater, and (x) temperature of
water at the exit of heat exchanger. 
All parameters are deduced from at least two redundant sensors except the residual gas pressures 
and the cooling water parameters.  The PLC continously evaluates the COPs, and determines the \GERDA \ safety status
in terms of familiar color codes. If some COPs deviate sigificantly 
from the foreseen range (see Table~\ref{tab:COPs}), it prompts automatically appropriate alarms and actions to mitigate the 
associated risks.
%%%%%%%%%%%%%%%%%%%%%%%%%%%%%%%%%%%%%%%%%%%%%%%%%%%%%%%%%%%%
\begin{table}[h]
\begin{center}
\caption{Critical operational parameters (COPs) and thresholds for mitigating actions. The threshold index {\it code} designates 
the category of the event which has triggered the respective alarm. 
}
\vskip+2truemm
\label{tab:COPs}
\begin{tabular} {lccccc}
\hline
\hline\\[-2.0ex]
\multicolumn{1}{l}{COP} & & \multicolumn{4}{c}{thresholds\,$_{code}$}\\[0.5ex]
\cline{3-6}\\[-2.0ex]  
                        &        &  green        & yellow      & orange &   red   \\[0.2ex]
\hline\\[-2.0ex]				 
{\bf Cryostat}         \\[0.2ex]
%\cline{1-1}
~$p$ [$10^5$\,Pa$_{\rm absolute}$]    & (i)       & $<1.5$        & $>1.5 $        &    \\             
~fill level [mm]         & (ii)       & $<-760$       & $>-760$        &    \\
~insulation vacuum                                                      \\
~~~$p$ [hPa]       & (iii)      & $<10^{-4}$    &                  & [$10^{-4}$, 0.1]\,$_{\rm o3}$ & $>$0.1\,$_{\rm r1}$ \\
~~~$p_{\rm res,air}$ [hPa]  & (iv)           & $<10^{-4}$    & $>10^{-4}$   &            \\
~~~$p_{\rm res,Ar}$ [hPa]   & (v)            & $<10^{-4}$    &            & $>10^{-4}$\,$_{\rm o2}$ \\
~~~$p_{\rm res,water}$ [hPa]& (vi)           & $<10^{-4}$    &             & $>10^{-4}$\,$_{\rm o3}$ \\[0.7ex]
{\bf Ar gas heater}     \\[0.2ex]     
%\cline{1-1}                        
~$\Delta p_{\rm Ar}$ [hPa]  & (vii)     & $<10$         &             & [10, 30]\,$_{\rm o4}$  & $>$30\,$_{\rm r2}$ \\
~$T_{\rm Ar}$ [$^\circ$C]     & (viii)     & $>2$          &             & [-5, 2]\,$_{\rm o5} $  & $<$-5\,$_{\rm r3}$\\
~$\dot m_{\rm water}$ [kg/s]& (ix)     & $>5$         & $<5$  \\
~$T_{\rm water}$  [$^\circ$C] & (x)     & $>6$          &             & [2, 6]\,$_{\rm o1}$     & $<$2\,$_{\rm r4}$\\[0.2ex]
\hline
\hline
\end{tabular}
\end{center}
\end{table}
%%%%%%%%%%%%%%%%%%%%%%%%%%%%%%%%%%%%%%%%%%%%%%%%%%%%%%%%%%%%%%%%%%%%%%%%%%%%%%%%%

\lq Green' indicates standard operation within the specifications. \lq Yellow' summarizes all events with no or marginal
impact on the safety in \hallA \ indicating the need for corrective maintenance. A typical event for this class 
would
be a loss of insulation vacuum due to an air leak. 
\lq Orange' designates  events with little immediate impact on the safety of work in \hallA \ but which might 
develop to an event of category \lq red'. A typical event would be a microscopic leak in one of the cryostat's 
walls.
\lq Red' designates the events of highest impact to the complete underground laboratory, e.g. a macrosopic
leak in one cryostat wall with the resulting high evaporation rate.
%The corresponding actions for the \lq orange' levels 1-3 include drainage of the water tank and 
%evacuation of \hallA .
The actions for the \lq orange' levels 1-3 (see Table~\ref{tab:COPs}) include drainage of the water 
tank, evacuation of \hallA , and for o4, o5 and all \lq red' events in addition increased 
ventilation in \hallA .

Since the filling of cryostat and water tank in 2012 no single red event has shown up. 
Two alarms in 2017 and 2019 have led to `orange' status.  
These will be discussed in section~\ref{subsec:performance}.    

%%%%%%%%%%%%%%%%%%%%%%%%%%%%%%%%%%%%%%%%%%%%%%%%%%%%%%%%%%%%%%%%%%%%%

%\input{CC201228-Comm.tex}

\section{Commissioning and Performance}

\subsection{Work at manufacturer}

Table~\ref{tab:acctest} shows a summary of the acceptance tests at the manufacturer site in chronological 
order. In addition, several Rn emanation measurements and cleaning cycles have been made that are discussed below
in section \ref{sec:workinhallA}.
%%%%%%%%%%%%%%%%%%%%%%%%%%%%%%%%%%%%%%%%%%%%%%%%%%%%%%%%%%%%
\begin{table}[h]
\begin{center}
\caption{Acceptance tests at manufacturer. Pressures quoted for the 
pressure tests are measured relative to atmospheric pressure. 
For Rn emanation tests see Table~\ref{tab:radonemanation}.
}
\vskip+2truemm
\label{tab:acctest}
\begin{tabular}{ll}
\hline
\hline \\[-2.0ex]
test & specification / result \\[0.5ex]
\hline \\[-2.0ex]
pressure test of inner vessel with water    & 3.6$\cdot 10^5$\,Pa \\
He leak test of inner vessel             & $<5\cdot 10^{-9}$\, hPa $\ell$ / s \\
load test of support pads                & passed \\
He leak test of outer vessel             & $<10^{-7}$\,hPa $\ell$ / s \\
\ln \ evaporation test                     & $<4$\,Nm$^3$/h or $<$300\,W \\
pressure test of outer vessel with N2       & 1.85$\cdot 10^5$\,Pa \\[0.5ex]
\hline
\hline
\end{tabular}
\end{center}
\end{table}
%%%%%%%%%%%%%%%%%%%%%%%%%%%%%%%%%%%%%%%%%%%%%%%%%%%%%%%%%%%
The maximum operating pressure of 1.5$\cdot 10^5$\,Pa relative to atmospheric pressure 
(2.5$\cdot 10^5$\,Pa differential with the insulation vacuum present) for the innner container 
implies a test pressure of 3.6$\cdot 10^5$\,Pa relative to atmospheric pressure. 
For the outer vessel the test pressure is 1.85$\cdot 10^5$\,Pa relative to atmospheric pressure. 
The pressure test of the inner container was done with 
water as pressurized medium.  
The pressure test of the outer container implies, however, the test of the fully assembled cryostat.
Hence it was done under appropriate safety precautions with nitrogen gas in order to prevent wetting of 
the superinsulation. 
The helium leak tests yielded the requested vacuum performance. Also the pump-down time of about 2 weeks
(including purging with dry nitrogen gas) for an insulation vacuum of about 10$^{-3}$\,hPa
turned out to be acceptable considering the additional polycarbonate layer below the MLI.    											

The inner vessel rests on 8 support pads. To ensure that all pads carry about equal load,
a special test was carried out: by temporarily mounting on each support pad a load cell, the weight on
each pad was measured after the empty inner vessel had been lowered onto the pads.
With each pad resting on a Belleville spring of known spring constant (58\,kN/mm), the differences between 
the measured loads translated into length differences of less than 0.12\,mm.   

\begin{figure} [h]
\begin{center}
\includegraphics[width=0.8\textwidth] {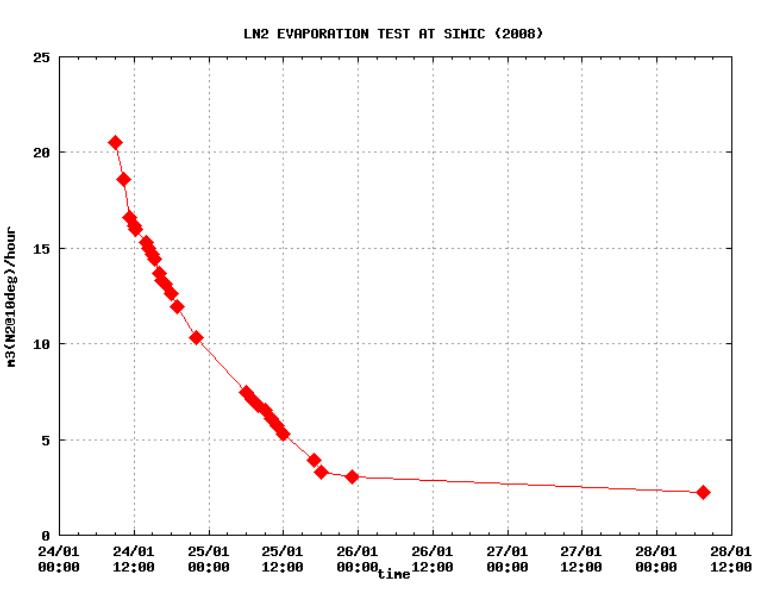}
\caption{Settling of evaporation rate after \ln \ fill.
}
\label{fig:evap}
\end{center}
\end{figure}

The evaporation rate of the cryostat has been determined with a fill of \ln \ up to the
neck at an insulation 
vacuum of 10$^{-3}$\,hPa. During cool down, temperatures were monitored at various locations of the cryostat by
sensors installed either below the MLI at the bottom and top of the vertical wall of the inner vessel
or mounted with appropriate thermal insulation on the vertical filling pipe inside the cryostat. 
\ln \ was distributed within the cryostat by a sprinkler with small holes covering 
a large surface and 
avoiding a cold spot. The flow of \ln \ was adjusted such that the cooling speed was less than 20~K/h 
and the maximum temperature difference among all sensors less than 50~K. A few hours after the completion 
of the fill, the spillway valve was closed, and the emerging gas was let via a bundle of ambient air heat 
exchanger columns to a GN2 flow meter. Fig.~\ref{fig:evap} shows the measured evaporation rate during
the settling time until equilibrium. The asymptotic flow of nitrogen is well below the specified rate of
4~m$^3$/h and corresponds to a thermal loss of less than 300\,W.   
%%%%%%%%%%%%%%%%%%%%%%%%%%%%%%%%%%%%%%%%%%%%%%%%%%%%%%%%%%%%%%%%%%%%%%%%%%%%%%%%%%%%%%%%%%%%%%%%
%%%%%%%%%%%%%%%%%%%%%%%%%%%%%%%%%%%%%%%%%%%%%%%%%%%%%%%%%%%%%%%%%%%%%%%%%%%%%%%%%%%%%%%%%%%%%%%%

\subsection{Work in \hallA \ of \LNGS}
\label{sec:workinhallA}

The cryostat was transported in horizontal orientation by flat bed truck to \LNGS . A metal rod inserted between
top flange and bottom of the inner vessel prevented any horizontal movement of the inner vessel. 
After arrival and erection in \hallA \ the evaporation rate of the cryostat was measured again, and the
previously measured result was reproduced.

\subsection*{Installation of internal copper shield}
The installation of the copper shield proceeded from a platform built around the cryostat's neck. 
A tent with an inside hoist was erected above the neck opening and kept at slight overpressure to 
limit the amount of dust entering the cryostat. The individual copper segments were brought with the hall  
crane up to the platform and transferred into the tent after removal of the the outer plastic cover. 
The segments were lowered with the hoist into the cryostat, freed there from the second plastic cover,
and picked up by a temporarily installed handling device that allowed to mount them at the
appropriate location. 

\subsection*{Radon emanation tests}
A crucial specification of the cryostat is its radon emanation. Monte Carlo studies indicate
that a homogeneous distribution of 8\,mBq of \rn222 \ within the inner vessel would add \vctsper \ to 
the background index. 

The emanation measurements were performed with the mobile extraction unit (MOREX) \cite{morex}. 
For each run the inner vessel was closed with a metal-sealed steel flange.  In order to remove
air-borne \rn222 \ the vessel was evacuated and filled twice with \rn222 -free nitrogen gas that
was purified during the fill with the cryogenic activated carbon traps of MOREX.  
The vessel was pressurized then up to 1.6$\cdot 10^5$\,Pa relative to atmospheric pressure
and subsequently closed to 
accumulate the emanating \rn222 . 
After 1 to 2 weeks the gas was mixed within the vessel to establish a homogeneous
\rn222 \ distribution, and at least two gas fractions of typical 20 to 40 m$^3$ were extracted
for the determination of their respective \rn222 \ concentration. Scaling these data to the total volume 
yielded then the total saturation emanation rates.       

Table~\ref{tab:radonemanation} shows the various emanation rates that have been determined at
various phases of the commissioning of the cryostat. 
%%%%%%%%%%%%%%%%%%%%%%%%%%%%%%%%%%%%%%%%%%%%%%%%%%%%%%%%%%%%
\begin{table}[h]
\begin{center}
\caption{$^{222}$Rn emanation in \GERDA \ cryostat. Quoted
results are the average of at least two extractions, uncertainties include 
both statistical and systematic contributions. 
% at different stages of commissioning
}
\vskip+2truemm
\label{tab:radonemanation}
\begin{tabular}{ccl}
\hline
\hline\\[-2.0ex]
 Test & Activity & Comment \\
      & [mBq]  &         \\[0.5ex]
\hline\\[-2.0ex]
1     & 23.3 $\pm$ 3.6 & cryostat after cleaning \\     % hmm frueher 23.4 +-6.5
%2     & 13.7 $\pm$ 1.9 & after additional cleaning \\
2a     & 13.6 $\pm$ 2.5 & after additional cleaning \\
2b     & 13.7 $\pm$ 2.8 &  \\
3     & 34.4 $\pm$ 6.0 & after Cu shield mounting \\
4     & 30.6 $\pm$ 2.4 & after wiping of inner surfaces \\
5     & 54.7 $\pm$ 3.5 & after mounting of infrastructure \\[0.5ex]
\hline
\hline
\end{tabular}
\end{center}
\end{table}
%%%%%%%%%%%%%%%%%%%%%%%%%%%%%%%%%%%%%%%%%%%%%%%%%%%%%%%%%%%
Two extractions after the first cleaning cycle of pickling, passivation and rinsing with de-ionized water
yielded rather different values, possibly due to a fractioning of Rn inside the vessel such that the
heavier Rn is concentrated at the bottom. Hence the Rn concentration is enhanced in the second sample. 
Moreover,
it was noted that the cryostat's inner surface was not everywhere metallic bright after the first cleaning.
Hence grinding at various spots and another cleaning cycle were done, and in the subsequent emanation
measurement the gas was mixed before the extraction of samples, one at the manufacturer's site and one after
transport to \LNGS . Two consistent and significantly lower values were measured which, considering the volume, 
compare very well with values measured for other selected cryogenic vessels. After the mounting of the internal
copper shield the emanation rate showed an unexpected large increase. Subsequent wiping of all
accessible steel and copper surfaces did barely improve the more precisely measured emanation value 
of about 31 mBq which is more than twice the value of the empty vessel. Later screening of the copper 
material revealed a surface contamination which presumably was not removed by the acid treatment at the 
rolling mill. The installation of the cryogenic infrastructure further increased the emanation level by
almost a factor of two. The measurements indicate that the new sources could be located in the 
pipes above the cryostat. 
 
  During data taking we did not detect a high radon level inside the
  liquid argon with the germanium detectors. This indicates that the
  radon emanating from the piping does not mix with the argon.

\subsection*{Final installation work}

%%%%%%%%%%%%%%%%%%%%%%%%%%%%%%%%%%%%%%%%%%%%%%%%%%%%%%%%%%%%%%%%%%%%%%%%%%%%%%%%%%%%%%%%%%%%%%%%
%\begin{figure} [h]
%\begin{center}
%\includegraphics[width=0.8\textwidth] {116002-Manifold_test_x.PNG}
%\caption{View from top through neck on active cooling system (), on the four CF200 ports connecting to the 
%insulation vacuum volume as well on the manifold mounted on top of the cryostat's with various feed throughs
%() () (). - >>> Figure not yet referenced in text. <<<
%}
%\label{fig:manifold1}
%\end{center}
%\end{figure}
%%%%%%%%%%%%%%%%%%%%%%%%%%%%%%%%%%%%%%%%%%%%%%%%%%%%%%%%%%%%%%%%%%%%%%%%%%%%%%%%%%%%%%%%%%%%%%%%

Work close to the cryostat included the construction of the surrounding water tank and 
the infrastructure building for \GERDA \ with the platform above the cryostat for cleanroom and lock. 
Latter activites caused the occasional deposition of carbon steel debris on some unprotected surfaces 
of the cryostat. 
When noticed, protection measures were improved, and after the end of the construction
work the surface of the cryostat was carefully inspected. All spots showing carbon steel
deposits were grinded, pickled and passivated. These activities as well as the installation of the
Jackodur thermal insulation took advantage of the scaffolding that was available within the
water tank for covering the inner walls of the water tank with the reflecting VM2000 foil.

With platform and cleanroom available, heat exchanger and radon shroud were inserted under
controlled clean conditions into the cryostat. With the manifold mounted the internal instrumentation
of the cryostat was connected to the outside cryogenic infrastructure including an ullage volume of 
2.6\,m$^3$. Connections of pipes were kept short by locating the latter close to the wall
of the water tanks with the exception of the \ln \ and LAr storage tanks.
The installation concluded by connecting the big DN630 shutter in the cleanroom via a double-walled 
bellow to the top of the manifold.

\subsection{Performance and operating experience }
\label{subsec:performance}

Initially, the insulation vacuum was pumped at a pressure drop of 1~-~2~Pa/s 
to avoid damage to the superinsulation.
Once the pressure was below 1 hPa a turbo pump was switched on and 
after two month of pumping a pressure of $10^{-5}$ hPa was reached
and the cryostat was filled. After cool down the pressure droped to
about $2\cdot 10^{-8}$ hPa. The turbo pump is running continously.
If it is switched off,  the 
vacuum pressure increases at
5$\cdot$10$^{-7}$\,hPa$\cdot\ell$/sec.

\subsection*{Inspection for corrosion after GERDA Phase I}
After more than 3 years of operation the water tank was drained in July 2013. The cryostat's outer thermal
insulation (Jackodur covered by VM2000 foil) appeared to be in perfect condition. Selected
welds and surfaces of water tank and cryostat were inspected by a representative of a notified body 
(INAIL, P.le Pastore 6, 00144 Roma).
No corrosion problems were observed (including surfaces below the thermal shield), and the safety 
of the pressure equipment was certified. Subsequently, the water tank was refilled and kept with two interruptions 
(see below) at its final level since September 2013.

\subsection*{Unscheduled drainage of water tank}
Two  alarms related to the insulation vacuum pressure
  triggered the drainage of the water tank during the 12 years of operation.
    In the first event (15/04/2017) one of the two pressure gauges was broken.
      The PLC recognized this condition and disabled the sensor for alarm
        generation. By human intervention, the faulty sensor was mistakenly
        enabled again and soon afterwards the alarm threshold was passed which
        triggered the drainage. 
	We decided to have shorter maintenance
	    intervals in order to reduce the risk of a sensor failure in the future.
	    
  The second event (11/11/2019) was triggered by a broken turbo pump such that the
    vacuum pressure increased within hours above the alarm threshold.
      The PLC closed the gate valve and separated the insulation vacuum from
        the pump. The cold cryostat wall then acted as a cryopump and the
	  insulation vacuum pressure quickly returned below the threshold.
	    The drainage however was started and had to be stopped manually.
	    
  We are discussing with \LNGS \ to modify the alarm and drainage conditions
    to reduce the risk of false alarms without jeopardizing the safety of
      the system.
               
Both events provided, however, evidence for the full functionality of the implemented safety system.

\subsection*{Drainage of LAr}
After completion of all  \GERDA \ measurements, the water tank and thereafter the cryostat were emptied.
The drainage of the LAr occured by evaporation. It started in September 2020 by increasing
the insulation vacuum pressure to about 1 hPa and later in steps to 5 hPa
and 30 hPa. Hence the heat transfer from the outside to the inner
steel vessel the LAr increased. We did not go higher in pressure since the expectation is that 
the heat transfer actually reduces at some point when going higher in pressure, and we also did 
not want to harm the MLI of the cryostat. The evaporation of the 90 tons of LAr took about 
7 weeks, longer than we expected from previous experience with \LN2 . 

The cryostat was filled twice with \LN2 \  at the time of production as part of the acceptance test. 
The evaporation rate at that time was much higher: emptying the cryostat took typically 2 weeks. 
We think the reason why it took much longer now is due to a) the larger heat to evaporate the
90 tons of argon compared to the 52 tons of nitrogen (40\%), 
b) the additional thermal insulation on the outside of the outer
cryostat wall (5 mm of extruded polystyrol foam and the VM1000 reflector foil),
and c) the reduced air volume inside the water tank.   

\subsection*{Inspection of the cryostat after GERDA Phase II}

In November 2020, i.e. 11 years after the LAr fill, an optical inspection of the cryostat has been 
performed both from the outside and inside.
No damage or deformation was observed. Moreover, cryostat and infrastructure passed without problem 
the required tests according to Italian regulations.

%%%%%%%%%%%%%%%%%%%%%%%%%%%%%%%%%%%%%%%%%%%%%%%%%%%%%%%%%%%%%%%%%%%%%%%%%%%%%%%%%%%%%%%%%%%%%%%%%%%%%%%%%%%%

\section{Conclusion}
 
The \GERDA \ experiment is the first realization of a novel shielding scheme using a cryostat filled with a liquid noble gas
that is operated in a large water tank, a concept that has since been adopted by other experiments like 
LUX \cite{LUX}, XENON1t \cite{XENON}, Darkside \cite{DARKSIDE}, or PandaX-4T \cite{PandaX}. 
The associated risks have been carefully analysed and mitigated. The major
construction materials of the cryostat have been all screened for radioactivity.
The cryogenic installation is running since December 2009 without any problem
that would have affected safety.
The background goals for \GERDA \ have been reached demonstrating the feasibility of a background-free
$0\nu\beta\beta$ decay experiment with $^{76}$Ge at the design exposure of 100~kg$\cdot$yr \cite{gphIIr4}.
%and have surpassed in Phase II, undercutting the design value for the background index of less 
%than 0.001 \ctsper \ in the region of the $0\nu\beta\beta$ decay line by a factor of 2. 
The experimental setup of \GERDA \ has been handed over in February 2020 to the \LEGEND \ collaboration \cite{LEGEND}
which is preparing the next generation search for \onbb \ decay of \gesix \ using thereby major parts
of the \GERDA \ experimental setup including water tank and cryostat with its infrastructure. 

%\section*{Acknowledgments}
\acknowledgments

The authors gratefully  acknowledge the generous and most helpful support by Ch.\,Haberstroh, H.\,Neumann 
and G.\,Perinic in numerous cryogenic issues. We acknowledge also the help
of V.N.\,Cryshtal and E.A.\,Bogdanov (Cryogenmash) in the design studies for a superinsulated
stainless steel cryostat with internal cold copper shielding. We thank our referees for critically
reading the manuscript and suggesting substantial improvements. 

\noindent
This work was supported by special funds of the Max Planck Society (MPG).

\clearpage
%\renewcommand{\arraystretch}{0.8}

%%%%%%%%%%%%%%%%%%%%%%%%%%%%%%%%%%%%%%%%%%%%%%%%%%%%%%%%%%%%%%%%%%%%%%%%%%%%%%
\appendix
\section{Heat transfer to LN2 and LAr in the pool boiling regime}

Studies of the gas exhaust rate for a worst case failure scenario of the \GERDA \ cryostat have been
performed on a small scale model.   
The experimental setup consisted of two concentric cylindrical containers of
about 50\,cm height, an inner one (\O 30\,cm) from stainless steel for the cryoliquid, and an outer one 
(\O 50\,cm) for the water. The inner container is resting on plastic feet for thermal insulation. 
The weight of both containers
including the reservoir containing the water is continously monitored with a digital scale whose display 
is quasi-continously read with a camcorder. Temperatures were measured with Pt100 sensors mounted within 1\,mm
at the outer surface of the wall of the inner container as well as at 10\,mm and 20\,mm distance from it.

%%%%%%%%%%%%%%%%%%%%%%%%%%%%%%%%%%%%%%%%%%%%%%%%%%%%%%%%%%%%%%%%%%%%%%%%%%%%%%%%%%%%%%%%%%%%%%%%
\begin{figure} [htb]
\begin{center}
\includegraphics[width=0.47\textwidth] {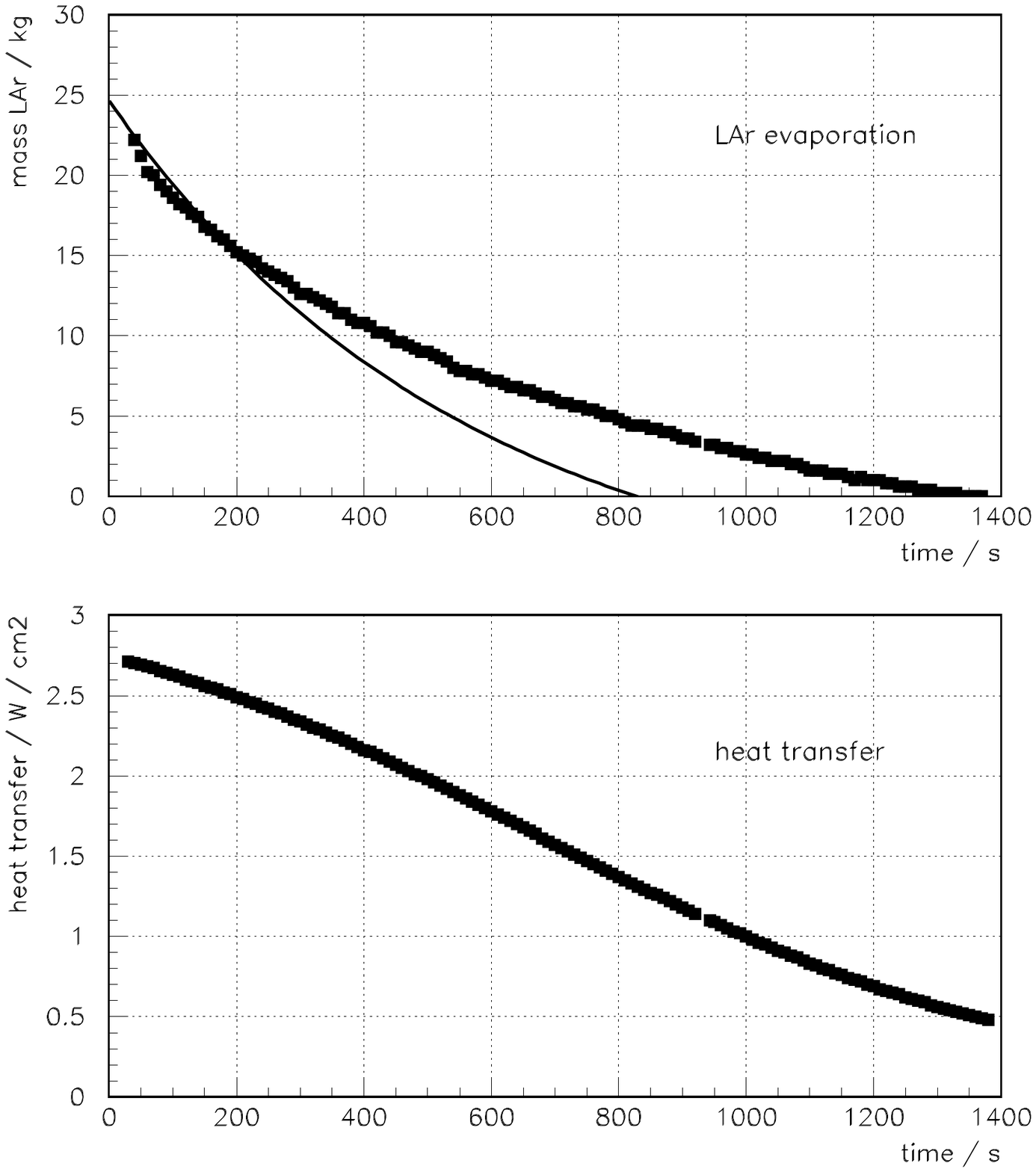}
\includegraphics[width=0.47\textwidth] {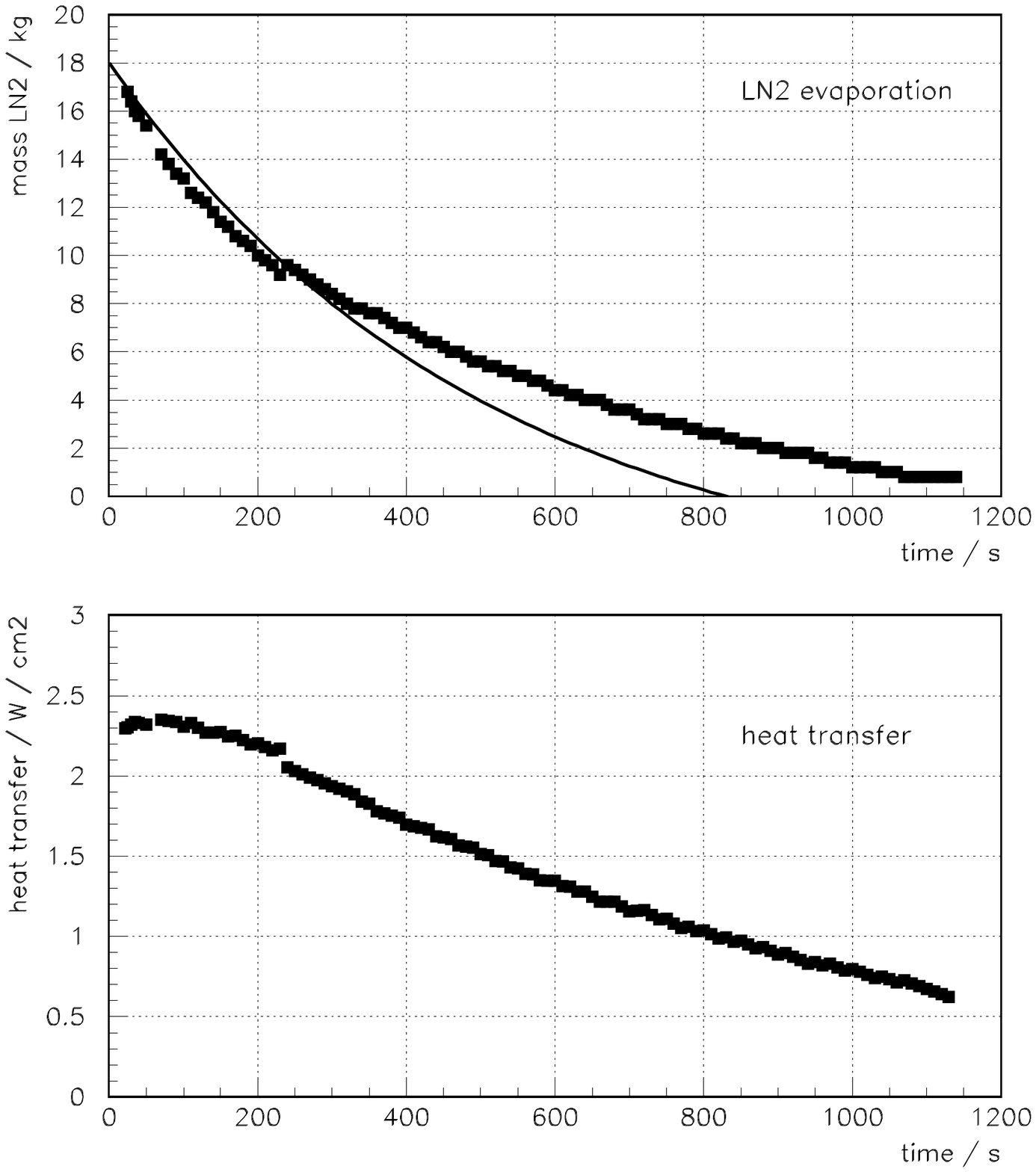} 
\caption{(top) Time dependence of the \ln \ and LAr mass after exposure of the container 
containing the cryoliquid to the water bath. The continuous lines are calculated with the 
integral of eq. (1) using a heat transfer coefficient of 24\,kW/m$^2$ for \ln , and 27\,kW/m$^2$ 
for LAr.  
(bottom) Time dependence of the heat transfer coefficients deduced with eq. (1) from the data 
shown on top.
}
\label{fig:evapln}
\end{center}
\end{figure}
%%%%%%%%%%%%%%%%%%%%%%%%%%%%%%%%%%%%%%%%%%%%%%%%%%%%%%%%%%%%%%%%%%%%%%%%%%%%%%%%%%%%%%%%%%%%%%%%

Fig.~\ref{fig:evapln} shows the results obtained for \ln \ and LAr. 
%together with deduced respective heat transfers. 
Data points obtained during the sudden filling and in the following 20 sec time interval are not 
shown since the setup was not adequate to map out these highly transient processes. 
Corresponding heat transfer coefficients $k$ were deduced using the relation 

$$ {{dM}\over{dt}} = -~{{k\cdot\pi\cdot R^2}\over {h}} -~{{2\cdot k}\over {\rho\cdot R\cdot 
h}}\cdot M~~~~~~~~~~(1)  $$

\noindent
where $M$, $\rho$ and $h$ denote mass, density and latent heat of the cryoliquid in the 
container of radius $R$ ($h$~=~162~kJ/kg and 199~kJ/kg for LAr and \ln , respectively).
The heat transfer coefficients decrease with time and we adopt as $k$ value the maximum of 
24~kW/m$^2$ and 27~kW/m$^2$ for \ln \ and LAr, respectively.    
The $k$ value for \ln \ is close to the expectation from the pool boiling characteristics 
of \ln \ \cite{Van63}. For LAr, we are not aware of published values.
The deduced $k$ values represent an upper limit since it is assumed that the surface covered by the 
cryogenic liquid is minimal at all times. This assumption neglects the well visible effect that 
the boiling liquid wells up the wall.

One reason for the decreasing heat transfer coefficients is the built up of an ice layer at the 
wall of the inner container. With a thermal conductivity of ice of about 3W/(m$\cdot$K) the 
corresponding heat transfer coefficient for a layer of 20\,mm thickness and $\Delta T$\,=\,100\,K 
is 15\,kW/m$^2$ limiting thus the effective heat flow. The temperature measurements performed 
with LAr show that the ice thickness of 10\,mm and 20\,mm is reached after about 300 and 600 
seconds.

%%%%%%%%%%%%%%%%%%%%%%%%%%%%%%%%%%%%%%%%%%%%%%5555

\section{Timeline of the \GERDA \ experiment}

\begin{table}[h]
\begin{center}
\caption{Timeline of the \GERDA \ experiment.
}
\vskip+2truemm
\label{tab:timeline}
\begin{tabular}{lc}
\hline
\hline \\[-2.0ex]
Event        & Date \\[0.5ex]
\hline\\[-2.0ex]
Letter of Intent to \LNGS           & Apr. 2004 \\
Proposal to \LNGS                   & Sep. 2004 \\
Technical Proposal                 & June 2006 \\
Risc Analysis finalized            & Feb. 2007 \\
Arrival of cryostat at \LNGS        & Mar. 2008 \\
Cryostat and water tank installed  & June 2008 \\
L'Aquila earthquake                 & Apr. 2009 \\
Cryostat filled with LAr           & Dec. 2009 \\
First detectors deployed in LAr    & June 2010 \\
Data taking Phase I                & Nov. 2011 - May 2013 \\
Upgrade to Phase II                & June 2013 - Nov.2015 \\ 
Data taking Phase II               & Dec. 2015 - May 2018 \\
Data taking Phase II               & July 2018 - Dec.2019 \\
\GERDA \ setup handed over to \legend        &  Feb. 2020 \\
\hline
\hline
\end{tabular}
\end{center}
\vskip-2mm
\end{table}

%\clearpage

%\input{CC201228-BIB.tex}

\clearpage

\end{document}